\documentstyle[12pt,aaspp4]{article}
\lefthead{JENKINS ET AL.}
\righthead{H2 TOWARD $\delta$ AND $\epsilon$ ORI}
\begin{document}
\title{The Properties of Molecular Hydrogen toward the Orion Belt Stars
from Observations by the Interstellar Medium Absorption Profile
Spectrograph}
\author{Edward B. Jenkins\altaffilmark{1}, Przemys{\l}aw R.
Wo\'zniak\altaffilmark{1}, Ulysses J. Sofia\altaffilmark{2,3}, George
Sonneborn\altaffilmark{4}, and Todd M. Tripp\altaffilmark{1}}
\altaffiltext{1}{Princeton University Observatory, Princeton, NJ 08544}
\altaffiltext{2}{Department of Astronomy and Astrophysics, 
Villanova University, Villanova, PA 19085}
\altaffiltext{3}{Present address: Dept. of Astronomy, Whitman College,
345 Boyer Ave., Walla Walla, WA 99362.}
\altaffiltext{4}{Code 681, NASA Goddard Space Flight Center, Greenbelt,
MD 20771}
\begin{abstract}
Absorption features from the Lyman and Werner bands of interstellar
molecular hydrogen were recorded by the Interstellar Medium Absorption
Profile Spectrograph (IMAPS) at $\lambda/\Delta\lambda=80,000$ in the
spectra of $\delta$~Ori~A and $\epsilon$~Ori.  The objective was to find
and study more examples of an unusual phenomenon found for one of the
velocity components of H$_2$ in the spectrum of $\zeta$~Ori by Jenkins
\& Peimbert  (1997).  Specifically, they detected a gradual shift in
velocity and broadening for features arising from progressively higher
rotational excitations $J$.  This effect appears to be absent in the
spectra of both $\delta$ and $\epsilon$~Ori, which are only a few
degrees away in the sky from $\zeta$~Ori.  The absence of atomic
material at a large negative velocity in the spectra of $\delta$ and
$\epsilon$~Ori (and its presence in $\zeta$~Ori) supports a proposal by
Jenkins \& Peimbert that the line of sight to $\zeta$ intercepts a bow
shock facing away from us, perhaps created by the collision of wind-like
material with some foreground obstruction.  One edge of the molecular
cloud complex Lynds~1630 is situated close to $\zeta$~Ori in the sky,
but we present some evidence that seems to indicate that the cloud is
more distant, in which case it could not serve as the obstruction. 
However it is possible that the outermost extension of a high-speed jet
from a star forming within the cloud can explain the high-velocity
material and the shock front created by it.  

For both stars, the H$_2$ absorption features are separated into two
velocity components.  Total H$_2$ column densities toward $\delta$ and
$\epsilon$~Ori are $5.5\times 10^{14}$ and $1.9\times 10^{16}{\rm
cm}^{-2}$, respectively.  When these values are compared to the column
densities of H~I, the fractions of H atoms bound in molecular form
$2N({\rm H}_2)/[2N({\rm H}_2)+N({\rm H~I})]= 7\times 10^{-6}$ for
$\delta$ and $1.3\times 10^{-4}$ for $\epsilon$.  The rotation
temperatures of the molecules with $J > 2$ toward $\epsilon$~Ori
indicate that the gas is in the general vicinity of the stars that emit
UV fluxes capable of rotationally pumping the molecules.  For the
strongest component of H$_2$ toward $\delta$~Ori, the pumping rate is
lower and consistent with a general UV flux level in the plane of the
Galaxy.
\end{abstract}
\keywords{ISM: Molecules --- stars: individual ($\delta$~Ori~A,
$\epsilon$~Ori, $\zeta$~Ori~A) --- Ultraviolet: ISM}

\section{Introduction}\label{intro}

The absorption bands of the Lyman and Werner systems arising from the
ground vibrational state of H$_2$ provide a rich array of electronic
transitions from different rotational levels.  It is often true that
these are the most conspicuous and plentiful features in the spectra of
early-type stars below about 1100\AA\  (Morton 1975; Jenkins et al.
1989; Snow, Allen, \& Polidan 1990) . In the interstellar medium, the
rotational excitation of molecular hydrogen is driven by a number of
processes, such as collisional excitation and de-excitation, optical
pumping by starlight, and the introduction of newly formed molecules. 
For this reason, the relative rotational populations of H$_2$ have been
used to gain insights on the characteristic molecule formation (and
destruction) rates, temperatures, and densities in the H$_2$-bearing
clouds  (Spitzer \& Cochran 1973; Spitzer et al. 1973; Spitzer, Cochran,
\& Hirshfeld 1974; Spitzer \& Zweibel 1974; Jura 1975a, b; Morton \&
Dinerstein 1976; Shull 1977, 1979; Shull \& York 1977).  The UV data
provide critical insights on the character of H$_2$ that are useful to
other kinds of observations and our current understanding of how various
environmental factors influence the rates of molecule formation and
destruction  (Shull \& Beckwith 1982).

In some cases, initial studies with the {\it Copernicus} satellite
revealed changes in the widths of the profiles for absorption features
arising from levels with different values of the rotational quantum
number $J$   (Spitzer \& Cochran 1973; Spitzer, Cochran, \& Hirshfeld
1974).  More exacting analyses of high quality recordings of the H$_2$
profiles in several stars by Spitzer \& Morton  (1976) indicated that
there were shifts in rotational temperatures for components with small
changes in radial velocity, as one might expect for the gas behind shock
fronts created by stellar winds.  More recently, a significant
improvement in the ability to study this phenomenon has come from the
much better wavelength resolving power of the Interstellar Medium
Absorption Profile Spectrograph (IMAPS)  (Jenkins et al. 1988), which
has flown on both sounding rockets and orbital missions  (Jenkins 1993,
1995; Jenkins et al. 1996).

Using data acquired by IMAPS on the ORFEUS-SPAS~I mission launched on
the STS-51 Space Shuttle flight in 1993, Jenkins \& Peimbert  (1997)
discovered that one of the velocity components of H$_2$ in front of
$\zeta$~Ori~A showed small, progressive velocity offsets and profile
broadenings as the rotation states increased from $J=0$ to 5.  They made
the conjecture that this behavior could arise from H$_2$ forming in the
collapsing column of partially ionized, cooling gas behind a nearly
stationary bow shock facing away from us.  This shock front was thought
to be created by the abrupt deceleration of wind-like material from the
star (or other stars nearby) as it collided with a dense cloud, perhaps
similar to ones seen elsewhere in Orion that have comet-like tails
facing away from prominent stars in the association  (Bally et al. 1991;
Cernicharo et al. 1992).  This interpretation was supported by
spectroscopic evidence for low-ionization material at $-94\,{\rm
km~s}^{-1}$ (interpreted as the high speed material feeding the shock on
its upstream side), and more highly ionized atoms at lower velocities
(immediate post-shock gas at $-36\,{\rm km~s}^{-1}$).  These velocities,
relative to the observed pileup of cold material at about $0\,{\rm
km~s}^{-1}$, implied a low compression ratio of 2.6.  Jenkins \&
Peimbert concluded that this outcome could be produced by an embedded
magnetic field in the pre-shock material that is oriented nearly
parallel to the flow.

The motivation of the present investigation of H$_2$ in front of the
other two bright stars in the belt of the Orion constellation,
$\delta$~Ori~A and $\epsilon$~Ori, was to gather independent evidence on
the phenomenon studied by Jenkins \& Peimbert  (1997) for $\zeta$~Ori~A. 
Since there are indications that the three stars are within a large
cavity containing x-ray emitting gas  (Burrows et al. 1993) and enclosed
by an expanding shell of cooler material  (Reynolds \& Ogden 1979)
powered by either stellar winds  (Castor, McCray, \& Weaver 1975; Weaver
et al. 1977) or by one or more explosive events  (Cowie, Songaila, \&
York 1979), it was not unreasonable to expect that we could witness
additional manifestations of the unusual behavior of H$_2$.  Tantalizing
indications that a phenomenon similar to that found by Jenkins \&
Peimbert was occurring on the line of sight to $\delta$~Ori~A were
discussed by Spitzer \& Morton  (1976).

The two additional Orion Belt targets were observed by IMAPS on its
second flight conducted in late 1996 aboard the ORFEUS-SPAS~II mission
on STS-80  (Hurwitz et al. 1998).  We present here the results of our
investigation of the absorption features arising from H$_2$.  The paper
is organized as follows.  The section on observations (\S\ref{obs})
starts with a description of some fundamental properties of the two
target stars (\S\ref{tgts}), $\delta$ and $\epsilon$~Ori, along with
$\zeta$~Ori~A which is featured as a contrasting case in later
discussions.  The IMAPS instrument and some properties of the data that
are relevant to our investigation are described in the next subsection
(\S\ref{inst}), and this is followed by a brief description of the
exposures that were taken (\S\ref{exposures}).  The section on analysis
(\S\ref{analysis}) describes some general steps that were taken in the
data reduction (\S\ref{general_steps}), outlines some special procedures
to register the velocities of features from H$_2$ in different $J$
levels (\S\ref{vel_regs}), shows the results of an investigation to
check on the relative $f$ values of the weakest lines out of $J=1$, 2
and 3 (\S\ref{validation}), and ends with a description of how we
constructed profiles of column density vs. velocity for each $J$ level. 
The fundamental qualitative properties of the H$_2$ features in the
spectra of $\delta$ and $\epsilon$~Ori are discussed in
\S\ref{qualitative}.  The distributions of H$_2$ over different $J$
levels are presented in \S\ref{distribution}, using the concept of a
rotation temperature over specific $J$ levels as a convenient way to
describe the relative populations.  We use these rotational temperatures
to synthesize a few approximate conclusions about the strengths of the
UV pumping fields.  In the section that describes our interpretation of
the results (\S\ref{interpret}), we note that the unusual shifts and
broadening of the H$_2$ profiles for different $J$ found by Jenkins \&
Peimbert  (1997) seem to be correlated with the presence (for
$\zeta$~Ori~A) or absence (for $\delta$ and $\epsilon$~Ori) of high
velocity gas (\S\ref{hi_vel}) or the possible proximity of $\zeta$~Ori
to a dense cloud with a high rate of star formation inside
(\S\ref{structures}).  Finally, in \S\ref{remarks} we present a few
remarks that pull together some of the concepts presented earlier and
discuss their possible meanings.

\section{Observations}\label{obs}

\subsection{Target Stars}\label{tgts}

The two stars studied in this paper, $\delta$~Ori~A and $\epsilon$~Ori,
have a separation in the sky of 1\fdg 44 and are members of the b
subgroup defined by Blaauw  (1964) within the Orion~OB1 association. 
Later in this paper, we will be making comparisons with data recorded on
an earlier IMAPS flight for $\zeta$~Ori~A and reported by Jenkins \&
Peimbert  (1997).  This star is nearly on the same line as that joining
$\delta$ and $\epsilon$, but it is 1\fdg 40 degrees from $\epsilon$ (see
Fig.~\ref{dust}).  For the benefit of readers who wish to compare the
H$_2$ results with other observations of interstellar material, the
equatorial\footnote{We have elected to use 1950 coordinates in
Figs.~\protect\ref{dust} and \protect\ref{gaustad} because nearly all of
the maps of interstellar structures in the current literature use this
coordinate system.} and galactic coordinates of the three stars are
given in Table~\ref{coord}.  Other important properties of these stars
are also given in the table.  All radial velocities reported in this
paper are heliocentric; the table lists the numerical values that must
be added to the heliocentric velocities to obtain those in the Local
Standard of Rest.

\placetable{coord}
\begin{deluxetable}{
l     
c     
c     
c     
}
\tablewidth{0pt}
\tablecaption{Coordinates and Other Properties of the Orion Belt
Stars\tablenotemark{a}\label{coord}}
\tablehead{
\colhead{Coordinate} &
\colhead{$\delta$~Ori~A} &
\colhead{$\epsilon$~Ori} &
\colhead{$\zeta$~Ori~A}\\
\colhead{or Property} &
\colhead{(HD 36486)} &
\colhead{(HD 37128)} &
\colhead{(HD 37742)} }
\startdata
$\alpha_{2000}$\dotfill&$5^{\rm h} 32^{\rm m} 0\fs 40$&$5^{\rm h}
36^{\rm m} 12\fs 81$&$5^{\rm h} 40^{\rm m} 45\fs 53$\nl
$\delta_{2000}$\dotfill&$-$0\arcdeg 17\arcmin 56\farcs 7&$-$1\arcdeg
12\arcmin 6\farcs 9&$-$1\arcdeg 56\arcmin 33\farcs 3\nl
$\alpha_{1950}\dotfill$&$5^{\rm h} 29^{\rm m} 27\fs 02$&$5^{\rm h}
33^{\rm m} 40\fs 48$&$5^{\rm h} 38^{\rm m} 14\fs 05$\nl
$\delta_{1950}$\dotfill&$-$0\arcdeg 20\arcmin 04\farcs 4&$-$1\arcdeg
13\arcmin 56\farcs 2&$-$1\arcdeg 58\arcmin 03\farcs 0\nl
$l$\dotfill&203\fdg 86&205\fdg 21&206\fdg 45\nl
$b$\dotfill&$-17\fdg74$&$-17\fdg24$&$-16\fdg59$\nl
$v_{\rm LSR}-v_{\sun}~({\rm
km~s}^{-1})$\dotfill&$-17.2$&$-17.4$&$-17.6$\nl
Spectral type\dotfill&B0$\,$III~+~O9$\,$V&B0$\,$Iae&O9.5$\,$Ibe\nl
$v\sin i$ (${\rm km~s}^{-1}$)\dotfill&144&91&140\nl
$V$\dotfill&2.23&1.70&2.05\nl
$E(B-V)$\dotfill&0.09&0.05&0.05\nl
$\log [N({\rm H~I})~({\rm
cm}^{-2})]$\dotfill&20.19\tablenotemark{b}&20.48\tablenotemark{c}&
20.39\tablenotemark{c}\nl
\enddata
\tablenotetext{a}{Stellar coordinates are from SIMBAD, while other
properties are from Hoffleit \& Jaschek  (1982) and Howarth, et al. 
(1997), except for $N$(H~I).}
\tablenotetext{b}{ (Jenkins et al. 1999).}
\tablenotetext{c}{ (Diplas \& Savage 1994).}
\end{deluxetable}

The Orion~OB1~b subgroup is at a distance modulus of approximately 7.8
(360~pc)  (Brown, de Geus, \& de Zeeuw 1994) to 8.0 (400~pc) (Warren \&
Hesser 1978).  The parallaxes measured by Hipparcos, $3.56\pm 0.83\,{\rm
mas}$ for $\delta$~Ori~A and $2.43\pm 0.91\,{\rm mas}$ for
$\epsilon$~Ori  (Perryman et al. 1997), are consistent with this range
of distances.  Both targets are bright, hot, and have a large projected
rotation velocity (see Table~\ref{coord}), making them excellent UV
light sources for viewing narrow interstellar absorption lines.

Since the instrument has no entrance slit, it is important to know if
there are companion stars which could introduce contamination signals to
the main spectra.  There are two companions to $\delta$~Ori~A with wide
separations from the visual primary.  One is a type B2V with a
separation of 52\arcsec\ (HD36485), but its visual magnitude is fainter
by 4.6, so its effect on the main spectrum is negligible.  The other at
33\arcsec\ (BD$-$00~983B) has $V=14$.  Two bright components of
$\delta$~Ori~A were identified by Heintz  (1980) and later confirmed by
the speckle interferometry observations  (McAlister \& Hendry 1982). 
Recent determinations of their separation yield a value 0\farcs267 
(Lindegren et al. 1997; Mason et al. 1998), which is much smaller than
the equivalent of our wavelength resolution projected onto the sky
(about 10\arcsec).  Finally, the visual primary is a partially
eclipsing, spectroscopic binary  (Koch \& Hrivnak 1981) with a period of
5.73~days  (Harvey et al. 1987).  Our other target, $\epsilon$~Ori, is a
spectroscopic binary with also a visual companion that is 3\arcmin\ away
and fainter by 8.7 magnitudes  (Hoffleit \& Jaschek 1982).

\subsection{Instrument}\label{inst}

Detailed descriptions of the IMAPS instrument and how it performs have
been presented by Jenkins et al.  (1988, 1996).  Briefly, IMAPS is an
objective-grating echelle spectrograph whose optical configuration is
very simple, consisting of a mechanical collimator (to reject the
off-axis light from stars other than the target), an echelle grating
with a 63\arcdeg\ blaze angle, a parabolic cross-disperser grating, and,
at the focal point, an electron-bombarded, windowless, intensified CCD
image sensor.  The nominal wavelength coverage of $930-1150$ \AA\ is
obtained in 4 sequential exposures that span the free spectral range of
the echelle format at different angles of grating tilt.  Spectra taken
at different angles were not of uniform quality because they had
different deviations off the optimum blaze angle of the echelle grating. 
This effect, combined with much lower reflection efficiencies of the
gratings below about 1000 \AA, resulted in large changes in signal
quality over different regions of the spectral format.  Nevertheless, we
were able to get good selections for lines out of various $J$ levels
(Tables~\ref{H2linelist_dori} and \ref{H2linelist_eori}), even after
defining a quality threshold for acceptance at a reasonably high level. 
The instrument is capable of attaining a wavelength resolution of order
$R\equiv\lambda/\Delta\lambda=2\times 10^5$, but on this mission an
out-of-focus condition resulted in $R\sim 8\times 10^4$ and some overlap
in energy between adjacent orders.  The effective area of IMAPS for
wavelengths greater than about 1000\AA\ (where most of the H$_2$
features that we used are located) is about $3\,{\rm cm}^2$.

There was a significant difference between the observations made on the
ORFEUS-SPAS~I and ORFEUS-SPAS~II flights, i.e., those reported by
Jenkins \& Peimbert  (1997) for $\zeta$~Ori~A and those given in this
paper, respectively.  On the earlier flight, there was inadequate
control of the humidity of the atmosphere inside the payload prior to
launch.  As a result, small water droplets formed on the hygroscopic
(and easily damaged) K$\,$Br photocathode during the decompression and
adiabatic cooling of the air inside the instrument that occurred during
the initial phases of the Shuttle launch.  As shown by Jenkins et al. 
(1996), all of the data recorded on the 1993 flight exhibited a blotchy
pattern of efficiency variations caused by the damage to the
photocathode.  Within each blotch, the quantum efficiency was about 50\%
of the peak sensitivity in the unaffected regions.  For the 1996 flight
of IMAPS, more stringent precautions were taken to control the
instrument's internal atmosphere prior to launch.  As a result, the
damage was prevented and photocathode response was very high, and
uniformly so over the entire format.

The photocathode damage on the first flight had both good and bad
consequences.  On the bad side, considerable effort during the reduction
of the data was needed to make the required corrections for the
variations.  In some places, these corrections may not have been
perfect, so unpredictable, systematic variations in intensity could
arise.  Also, the overall efficiency of the detector was somewhat less
than it would have been otherwise.  On the good side, the damage to the
photocathode had a beneficial effect of providing an accurate indication
of the detector's geometrical mapping function.  By correlating the
apparent positions of the blotches in the data with a post-flight
photograph of the damaged photocathode, accurate corrections for the
distortion of the recorded spectral image could be implemented during
data processing, with the end result of producing accurate wavelength
determinations for all of the absorption features.  

For the data from the second flight reported on here, we had excellent
looking raw data frames but no information about the geometric
distortions (aside from the bending of the orders -- information which
is of no use in understanding perturbations in the wavelength scale). 
It was clear that we could not use the mapping function from the
previous flight for the $\delta$ and $\epsilon$~Ori data because there
was a change in the stray magnetic fields from other systems (which
affect the electron trajectories in the detector) from the first flight
to the second.  Thus, in short, for the $\zeta$~Ori data extraordinary
measures were needed to correct for photometric errors, while wavelength
derivations were relatively easy.  For the $\delta$ and $\epsilon$~Ori
data the reverse was true: obtaining relative photometric information of
good quality was straightforward, but special efforts were needed
(described in \S\ref{vel_regs}) to obtain accurate radial velocities.

\subsection{Exposures}\label{exposures}

\begin{figure}
\plotone{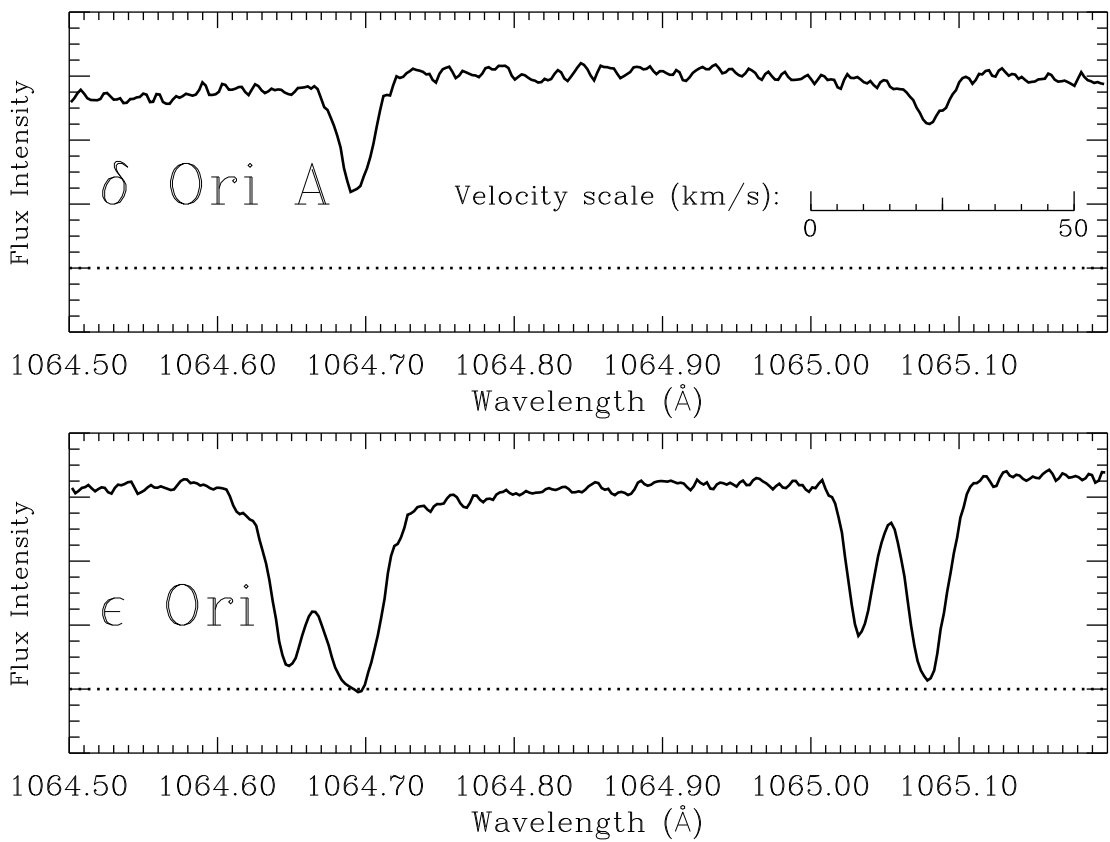}
\caption{Small samples of the spectra of $\delta$~Ori~A and
$\epsilon$~Ori recorded by IMAPS.  The left-hand feature is the Lyman
Band 3$-$0~P(1) line while the one on the right is the 3$-$0~R(2) line. 
The profiles of H$_2$ absorption are double peaked in the spectrum of
$\epsilon$~Ori.  A representative scale for radial velocities is shown
in the upper panel.\label{h2sample}}
\end{figure}

The exposures for $\delta$~Ori~A with a total integration time of 6064~s
were spread over 3 sessions: 1996 November 22 01:39--02:15~UT, November
29 12:13--12:54~UT, and December 3 21:33--22:12~UT.  The last session
had all exposures taken at a single grating tilt to enhance the quality
of the spectra at the very short wavelengths of the deuterium L$\delta$
and L$\epsilon$ transitions, used ultimately for the study of atomic D/H 
(Jenkins et al. 1999).  An additional set of exposures was taken when
the spacecraft was in a minimum drag configuration [to avoid a possible
collision with another satellite  (Hurwitz et al. 1998)], but the
unusual aspect of the Sun created an unexpected change in thermal
conditions that shifted the cross-disperser coverage to much longer
wavelengths, ones that were inappropriate for studying the H$_2$
transitions.  For $\epsilon$~Ori, a contiguous set of exposures was
taken during a single orbit over the time interval November 29
04:33--05:20~UT, for a total exposure time of 2452~s.  Sample portions
of the spectra covering two H$_2$ absorption features are shown in
Fig.~\ref{h2sample}.

\placefigure{h2sample}

\section{Analysis}\label{analysis}
\subsection{General Steps}\label{general_steps}

To derive an approximate wavelength scale, dispersion constants for the
various exposures were derived from the observed positions of H$_2$
features.  However, we did not assume that lines out of different $J$
levels or with very different strengths necessarily had the same
apparent radial velocity.  This analysis could be carried out because
the H$_2$ lines are so plentiful (we used approximately twice as many
lines as those selected for analysis and listed in
Tables~\ref{H2linelist_dori} and \ref{H2linelist_eori}).  We were able
to create outcomes that were internally consistent to within about
$0.87\,{\rm km~s}^{-1}$ for $\delta$~Ori~A and $0.62\,{\rm km~s}^{-1}$
for $\epsilon$~Ori, as judged by the rms dispersion of apparent residual
wavelength offsets for lines coming from any given $J$ level.  The lower
rms dispersion of $0.5\,{\rm km~s}^{-1}$ obtained by Jenkins \& Peimbert 
(1997) reflected the better control they had over their wavelength
scales, as discussed in \S\ref{inst}.  An important objective of our
investigation was to detect any shifts in velocity across different $J$
levels.  For this purpose, we devised a more precise tuning of the
wavelengths for just the H$_2$ profiles, so that we could register them
with better accuracy.  This procedure is described in \S\ref{vel_regs}
below.

Before extracting the spectra from the echelle orders, a distortion
correction procedure was invoked to remap the images so that the orders
were approximately straight and horizontal.  Even though this was done,
we found it necessary for the extraction program to determine the exact
small residual vertical shifts (in the cross-dispersion direction) for
each order as a function of horizontal position (along the echelle
dispersion direction).

The distribution of energy across each order in the vertical direction
varied with position, owing to small changes in aberrations in the
image.  Thus, in addition to characterizing the changes in the centers
of the orders, we also measured their widths and how they varied with
position. After the orders had been characterized, we performed an
optimum extraction of spectral information using the formulae given by
Jenkins et al.  (1996).  This method not only gives the best
signal-to-noise ratio, but it also compensates for the small spillover
of energy between adjacent orders.  This is needed to prevent the
appearance of spurious, ghost-like features that would ordinarily arise
if a simple ``dumb-slit'' extraction were performed.

For deriving the scattered light background levels, we obtained guidance
from saturated atomic lines with flat bottoms.  We also verified that
our equivalent widths agreed reasonably well with those measured by
Spitzer, Cochran \& Hirshfeld  (1974).  In some cases, we had to make
adjustments of up to 20\% in the background levels to make some lines
agree with others.

\subsection{Accurate Velocity Registration of H$_2$
Lines}\label{vel_regs}

To reduce the errors in radial velocities caused by image distortions,
we found pairs of H$_2$ features that were near each other on the
echelle spectrum format, but which came from different $J$ levels. 
Lines on a single order that were separated by about 0.1\AA\ (or less)
or lines on adjacent orders (or nearly so) with a very small horizontal
separation were deemed to be suitable for this purpose.  The identities
of these combinations are given in the endnotes of
Tables~\ref{H2linelist_dori} and \ref{H2linelist_eori}.  We then
measured the differences in radial velocity and used them to define the
relative shifts of the absorptions from different $J$ levels with
respect to each other.  The shifts averaged out to nearly $0\,{\rm
km~s}^{-1}$ for the features in both $\delta$ and $\epsilon$~Ori, with
the largest single deviation being about $1\,{\rm km~s}^{-1}$.  In using
the line pairs to define the velocity offsets, our working assumption
was that image distortions over very small scales were absent, and thus
the offsets were identical for both lines.

To determine an absolute reference of the velocity scale, we compared
the positions of the 5$-$0~R(0) and 5$-$0~R(1) lines to three prominent
telluric features of O~I out of excited fine-structure levels.  These
lines are rarely seen to arise from interstellar clouds, and thus they
serve as excellent velocity calibration features in our spectra.

\subsection{A Validation of $f$-values for some Weak
Lines}\label{validation}

Jenkins \& Peimbert  (1997) encountered some difficulty in obtaining
consistent results for different combinations of lines of H$_2$ arising
from the weakest members of the Lyman bands out of $J=0$, 1 and 2.  It
was possible that these inconsistencies could have arisen from either
uncorrectable saturation effects, systematic errors in the background
levels, errors in the adopted $f$-values, or some combination of the
three.  The amount of H$_2$ in the direction of $\delta$~Ori~A is not
very high.  For this reason, the weak H$_2$ transitions should have very
small corrections for saturation, which leads to an opportunity to check
on the reasonableness in the relative $f$-values.  This is an important
check to make, since these lines provide the only means for us to
measure the strongest parts of the features out of $J=0$, 1 and 2 for
$\epsilon$~Ori.

\begin{figure}
\plotone{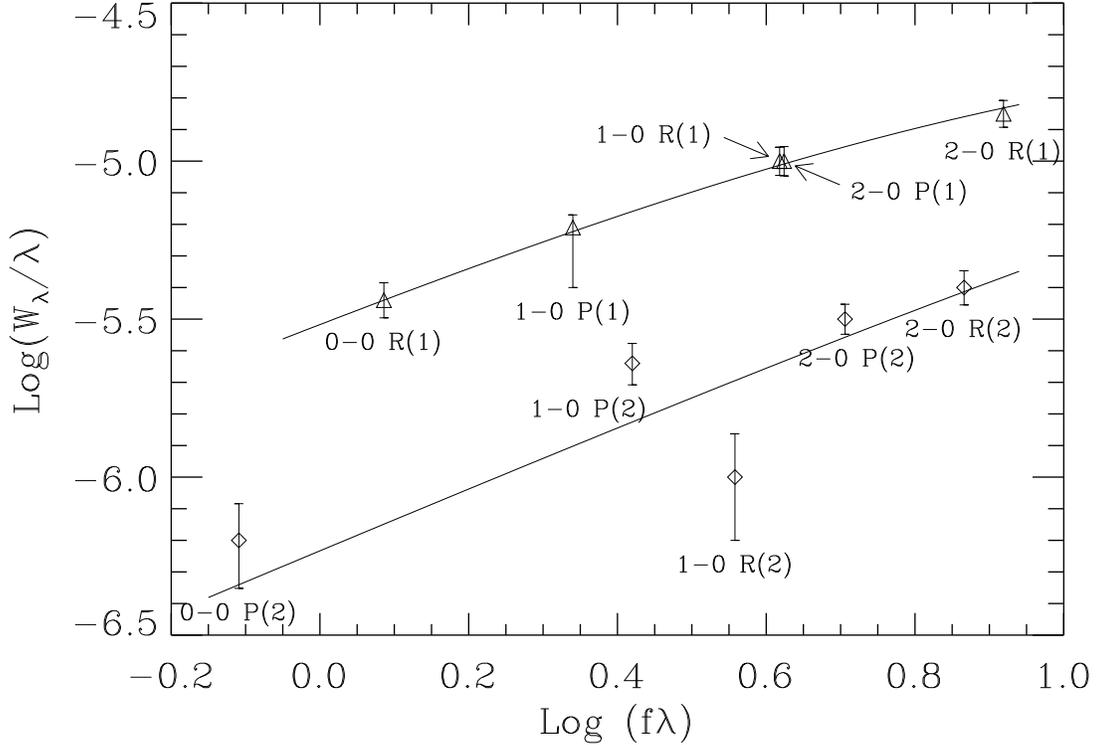}
\caption{The observed strengths of weak H$_2$ features out of $J=1$
(triangles) and 2 (diamonds), expressed in terms of
$\log(W_\lambda/\lambda)$ in the spectrum of $\delta$~Ori~A, against the
published values of $\log(f\lambda)$  (Abgrall \& Roueff 1989).  The
error bars show the $1\sigma$ uncertainties caused by noise and a 10\%
error in the background estimate [except for the recording of the
1-0~P(1) line, where the background uncertainty was larger].  The curves
show the calculated shapes of the curve of growth for the lines, based
on Na~I absorptions observed by Welty et al.  (1994).  The horizontal
placement of these curves are shifted to best match the points for $J=1$
and 2.\label{weaklins}}
\end{figure}

Figure~\ref{weaklins} shows the curves of growth for weakest lines out
of $J=1$ and 2 in our spectrum of $\delta$~Ori~A.  Unfortunately, the
weakest lines out of $J=0$ were too weak to give meaningful information. 
To estimate the amount of possible saturation in the H$_2$ lines, we
computed their expected behavior for a velocity profile that was
identical to the superposition of components derived for Na~I in front
$\delta$~Ori~A by Welty, Hobbs \& Kulkarni  (1994), based on their much
higher resolution recordings of the D1 line. (There is no reason to
expect the H$_2$ to exactly mimic Na~I, but this is the best example
that we can draw upon.  We do not want to use our own H$_2$ data because
there may be very narrow features that are not evident at our
resolution.)  The reasonably good correspondence between our
measurements and the curves in the figure indicates that the adopted
relative $f$-values from the theoretical calculations of Abgrall \&
Roueff  (1989) are reasonably accurate.

\placefigure{weaklins}

\subsection{Column Densities vs. Velocities}\label{col_dens}

After correcting individual line velocity offsets (\S\ref{vel_regs}), we
derived the apparent differential column densities per unit velocity for
each transition given by the expression
\begin{equation}\label{N_a}
N_a(v) = 3.768\times 10^{14}~{\tau_a(v)\over f\lambda}~{\rm
cm}^{-2}({\rm km~s}^{-1})^{-1}~,
\end{equation}
where $f$ is the transition's $f$-value and $\lambda$ is expressed in
\AA.  The apparent absorption optical depth $\tau_a(v)$ is defined in
terms of the continuum level $I_0$ and measured intensity $I(v)$ by the
relation
\begin{equation}\label{tau_a}
\tau_a(v) = \ln \left( {I_0\over I(v)}\right) ~.
\end{equation}
For each $J$ level, we consolidated the $N_a(v)$'s for different lines
into a single composite profile with the technique devised by Jenkins \&
Peimbert  (1997).  Not all of the lines that were visible in the
spectrum were included in the composite.  Some were rejected because
they had interference from other lines, were too close to the edge of an
image, or fell below a threshold that we defined for the minimum
acceptable signal-to-noise ratio.  The lines that were chosen (except
where noted) are listed in Tables~\ref{H2linelist_dori} and
\ref{H2linelist_eori}.

For $\epsilon$~Ori, even the weakest lines out of $J=0$, 1 and 2 were
saturated.  In the central portions of these lines, we implemented the
correction technique developed by Jenkins  (1996) to compensate for the
error between the apparent optical depth $\tau_a(v)$ and a smoothed
representation of the proper optical depth $\tau(v)$.  The pairs of
lines that were used in each case are identified in
Table~\ref{H2linelist_eori}.  The other lines for these three $J$ levels
were used to define the shoulders of the $N_a(v)$ distributions.

The usefulness of the correction technique is strongly driven by how
well one knows the zero intensity level.  If it is poorly known, there
can be an ambiguity in the interpretation of lines that do not have
$\tau_a(v)$ scaling in proportion to $f\lambda$.  A poor scaling can
arise either if the background levels are wrong or if the lines have
unresolved, saturated substructures.  Unfortunately, there are no strong
atomic lines in the vicinity of the weakest H$_2$ lines.  Thus, for
these cases we must acknowledge the existence of background
uncertainties and their possible influence on the accuracy of the
corrections for saturated absorptions for $J=0$, 1 or 2 in
$\epsilon$~Ori.  The increased errors are reflected in column~5 of
Table~\ref{Ns_epsilon}. 

Lines out of $J > 2$ for $\epsilon$~Ori and all of the H$_2$ lines in
the spectrum of $\delta$~Ori~A either showed good consistency in their
$N_a(v)$ profiles from strong to weak lines, or they were so weak that
the distortions caused by saturated sub-structures should not be a
problem.  Also, the background levels for these lines are probably more
reliable than for those that we had to use for $J=0$, 1 and 2 for
$\epsilon$~Ori because there were nearby, saturated atomic lines.

\placetable{H2linelist_dori}
\placetable{H2linelist_eori}
\begin{deluxetable}{
r     
r     
r     
r     
}
\tablecolumns{4}
\footnotesize 
\tablewidth{400pt}
\tablecaption{H$_2$ Lines for $\delta$~Ori~A\label{H2linelist_dori}}
\tablehead{
\colhead{Ident.\tablenotemark{a}} & 
\colhead{$\lambda$ (\AA)} &
\colhead{Log ($f\lambda$)} &
\colhead{S/N} }
\startdata
\cutinhead{$J=0$}
1$-$0 R(0)\tablenotemark{b}&1092.195&0.802&89\nl
4$-$0 R(0)\tablenotemark{c}&1049.367&1.383&19, 36\nl
5$-$0 R(0)\tablenotemark{c,d,e}&1036.545&1.447&46, 78\nl
7$-$0 R(0)\tablenotemark{c,f}&1012.810&1.483&26, 36\nl
10$-$0 R(0)\phm{\/}&981.437&1.314&17\nl
W 0$-$0 R(0)\tablenotemark{g}&1008.552&1.647&22\nl
\cutinhead{$J=1$}
0$-$0 R(1)\phm{\/}&1108.632&0.086&35\nl
1$-$0 P(1)\phm{\/}&1094.052&0.340&34\nl
1$-$0 R(1)\phm{\/}&1092.732&0.618&32\nl
2$-$0 P(1)\tablenotemark{c}&1078.927&0.624&19, 25\nl
2$-$0 R(1)\tablenotemark{c,h}&1077.700&0.919&18, 35\nl
3$-$0 P(1)\phm{\/}&1064.606&0.805&22\nl
3$-$0 R(1)\tablenotemark{c}&1063.460&1.106&15, 16\nl
4$-$0 P(1)\tablenotemark{e,i,j}&1051.033&0.902&25, 51, 71\nl
4$-$0 R(1)\tablenotemark{c}&1049.960&1.225&24, 34\nl
5$-$0 P(1)\tablenotemark{c}&1038.157&0.956&23, 39\nl
5$-$0 R(1)\tablenotemark{c,d}&1037.149&1.271&42, 76\nl
7$-$0 P(1)\tablenotemark{c}&1014.325&0.960&14, 21\nl
7$-$0 R(1)\tablenotemark{i}&1013.434&1.307&26, 36, 70\nl
8$-$0 P(1)\tablenotemark{c}&1003.294&0.931&21, 23\nl
8$-$0 R(1)\phm{\/}&1002.449&1.256&18\nl
10$-$0 P(1)\tablenotemark{i}&982.834&0.825&14, 21, 36\nl
10$-$0 R(1)\phm{\/}&982.072&1.138&17\nl
W 0$-$0 Q(1)\phm{\/}&1009.770&1.384&16\nl
W 0$-$0 R(1)\tablenotemark{g}&1008.498&1.326&21\nl
W 1$-$0 Q(1)\tablenotemark{c}&986.796&1.551&14, 29\nl
W 2$-$0 Q(1)\tablenotemark{i}&966.093&1.529&14, 16, 23\tablebreak
\cutinhead{$J=2$}
1$-$0 P(2)\phm{\/}&1096.438&0.420&28\nl
2$-$0 P(2)\tablenotemark{c}&1081.266&0.706&53, 95\nl
2$-$0 R(2)\phm{\/}&1079.226&0.866&32\nl
3$-$0 R(2)\tablenotemark{c}&1064.994&1.069&25, 31\nl
4$-$0 R(2)\tablenotemark{i}&1051.498&1.168&30, 40, 80\nl
5$-$0 P(2)\tablenotemark{c}&1040.366&1.017&28, 41\nl
5$-$0 R(2)\phm{\/}&1038.689&1.221&44\nl
7$-$0 P(2)\phm{\/}&1016.458&1.007&31\nl
8$-$0 R(2)\phm{\/}&1003.982&1.232&35\nl
12$-$0 P(2)\phm{\/}&966.273&0.798&31\nl
W 0$-$0 P(2)\phm{\/}&1012.169&0.746&29\nl
W 0$-$0 R(2)\tablenotemark{i}&1009.024&1.208&26, 29, 58\nl
W 2$-$0 R(2)\phm{\/}&965.791&1.490&25\tablebreak
\cutinhead{$J=3$}
2$-$0 R(3)\tablenotemark{c,b}&1081.712&0.840&52, 71\nl
3$-$0 P(3)\tablenotemark{c}&1070.141&0.910&23, 33\nl
3$-$0 R(3)\tablenotemark{h}&1067.479&1.028&28\nl
4$-$0 P(3)\tablenotemark{i}&1056.472&1.006&35, 42, 69\nl
4$-$0 R(3)\phm{\/}&1053.976&1.137&27\nl
5$-$0 P(3)\phm{\/}&1043.502&1.060&29\nl
5$-$0 R(3)\tablenotemark{c}&1041.157&1.222&31, 103\nl
7$-$0 R(3)\tablenotemark{c}&1017.422&1.263&30, 36\nl
8$-$0 P(3)\tablenotemark{g,k}&1008.383&1.004&21\nl
9$-$0 R(3)\tablenotemark{c}&995.970&1.229&24, 58\nl
11$-$0 P(3)\phm{\/}&978.217&0.817&32\nl
W 0$-$0 Q(3)\tablenotemark{c,f}&1012.680&1.386&24, 40\nl
W 1$-$0 P(3)\phm{\/}&991.378&1.075&39\nl
\enddata
\tablenotetext{a}{All transitions are in the
2p$\sigma\,B\,^1\Sigma_u^+\leftarrow {\rm X}\,^1\Sigma_g^+$ Lyman band
system, unless preceded with a ``W'' which refers to the
2p$\pi\,C\,^1\Pi_u\leftarrow {\rm X}\,^1\Sigma_g^+$ Werner bands.}
\tablenotetext{b}{The 1$-$0~R(0) line is very near the 2$-$0~R(3) line 2
echelle orders away.  These two lines were used to register lines out of
$J=0$ to those from $J=3$, as described in \S\protect\ref{vel_regs}.}
\tablenotetext{c}{This line was observed twice.}
\tablenotetext{d}{The velocity scale for this line was forced to be
consistent with the placement of the telluric O~I* (1027.431 and
1040.943\AA) and O~I** (1041.688\AA) lines at their computed
heliocentric velocities at the time of observation.  See
\S\protect\ref{vel_regs} for details.}
\tablenotetext{e}{The 5$-$0~R(0) line is very near the 4$-$0~P(1) line 3
echelle orders away.  These two lines were used to register lines out of
$J$=0 to those from $J=1$, as described in \S\protect\ref{vel_regs}.}
\tablenotetext{f}{The proximity of the 7$-$0~R(0) and W~0$-$0~Q(3) lines
to each other ($\Delta\lambda=0.13$\AA) enables an accurate registration
of the velocities of the lines out of $J=0$ to those out of $J=3$.}
\tablenotetext{g}{The 8$-$0~P(3), W~0$-$0~R(1) and W~0$-$0~R(0) are
separated by 0.11 and 0.05\AA, providing an opportunity to register the
velocity scales for the lines out of $J=0$, 1 and 3.}
\tablenotetext{h}{The 2$-$0~R(1) line is very near the 3$-$0~R(3) line 2
orders away.  These two lines were used to register lines out of $J=1$
to those from $J=3$, as described in \S\protect\ref{vel_regs}.}
\tablenotetext{i}{This line was observed three times.}
\tablenotetext{j}{Not used in building a composite profile.}
\end{deluxetable}
\begin{deluxetable}{
r     
r     
r     
r     
}
\tablecolumns{4}
\footnotesize 
\tablewidth{400pt}
\tablecaption{H$_2$ Lines for $\epsilon$~Ori\label{H2linelist_eori}}
\tablehead{
\colhead{Ident.\tablenotemark{a}} & 
\colhead{$\lambda$ (\AA)} &
\colhead{Log ($f\lambda$)} &
\colhead{S/N} }
\startdata
\cutinhead{$J=0$}
0$-$0 R(0)\tablenotemark{b}&1108.127&0.275&34\nl
1$-$0 R(0)\tablenotemark{b,c}&1092.195&0.802&54\nl
4$-$0 R(0)\tablenotemark{d}&1049.367&1.383&32\nl
5$-$0 R(0)\tablenotemark{e,f}&1036.545&1.447&46\nl
7$-$0 R(0)\tablenotemark{g}&1012.810&1.483&34\nl
\cutinhead{$J=1$}
0$-$0 P(1)\tablenotemark{h,i}&1110.063&$-$0.191&11\nl
0$-$0 R(1)\tablenotemark{b}&1108.632&0.086&41\nl
2$-$0 P(1)\tablenotemark{b}&1078.927&0.624&29\nl
2$-$0 R(1)\tablenotemark{i,j}&1077.700&0.919&18.1\nl
3$-$0 R(1)\phm{\/}&1063.460&1.106&30\nl
3$-$0 P(1)\phm{\/}&1064.606&0.805&46\nl
4$-$0 P(1)\tablenotemark{f}&1051.033&0.902&47\nl
4$-$0 R(1)\phm{\/}&1049.960&1.225&39\nl
5$-$0 R(1)\tablenotemark{e}&1037.149&1.271&55\nl
7$-$0 R(1)\phm{\/}&1013.434&1.307&45\nl
8$-$0 P(1)\phm{\/}&1003.294&0.931&26\nl
W 2$-$0 Q(1)\phm{\/}&966.093&1.529&17\nl
\cutinhead{$J=2$}
0$-$0 R(2)\tablenotemark{h}&1110.120&0.018&11\nl
1$-$0 P(2)\tablenotemark{b}&1096.438&0.420&43\nl
2$-$0 P(2)\tablenotemark{b}&1081.266&0.706&53\nl
2$-$0 R(2)\phm{\/}&1079.226&0.866&34\nl
3$-$0 R(2)\phm{\/}&1064.993&1.069&47\nl
4$-$0 P(2)\phm{\/}&1053.284&0.982&22\nl
4$-$0 R(2)\phm{\/}&1051.498&1.168&48\nl
5$-$0 P(2)\phm{\/}&1040.366&1.017&44\nl
5$-$0 R(2)\tablenotemark{k}&1038.689&1.221&20\nl
7$-$0 P(2)\phm{\/}&1016.458&1.007&26\nl
8$-$0 R(2)\phm{\/}&1003.982&1.232&37\nl
9$-$0 P(2)\phm{\/}&994.871&0.937&15\nl
W 0$-$0 P(2)\phm{\/}&1012.169&0.746&30\nl
W 0$-$0 R(2)\phm{\/}&1009.024&1.208&33\nl 
W 0$-$0 Q(2)\tablenotemark{l}&1010.938&1.385&18\nl
\cutinhead{$J=3$}
1$-$0 R(3)\tablenotemark{m}&1096.725&0.531&45, 45\nl
2$-$0 R(3)\tablenotemark{c}&1081.712&0.840&52\nl
4$-$0 P(3)\phm{\/}&1056.472&1.006&48\nl
3$-$0 R(3)\tablenotemark{i,j}&1067.479&1.028&20.5\nl
4$-$0 R(3)\phm{\/}&1053.976&1.137&34\nl
5$-$0 P(3)\tablenotemark{i,k}&1043.502&1.060&22.1\nl
5$-$0 R(3)\phm{\/}&1041.157&1.222&42\nl
6$-$0 P(3)\phm{\/}&1031.192&1.055&44\nl
7$-$0 R(3)\phm{\/}&1017.422&1.263&27\nl
8$-$0 R(3)\tablenotemark{i,l}&1006.411&1.207&9.5\nl
9$-$0 R(3)\phm{\/}&995.970&1.229&34\nl
W 0$-$0 Q(3)\tablenotemark{g}&1012.680&1.386&28\nl
\cutinhead{$J=4$}
3$-$0 P(4)\phm{\/}&1074.313&0.923&44\nl
3$-$0 R(4)\phm{\/}&1070.900&1.012&43\nl
4$-$0 P(4)\phm{\/}&1060.581&1.017&33\nl
5$-$0 R(4)\tablenotemark{d}&1044.542&1.195&32\nl
6$-$0 R(4)\phm{\/}&1032.349&0.814&22\nl
9$-$0 R(4)\phm{\/}&999.268&1.204&24\tablebreak
\cutinhead{$J=5$}
4$-$0 R(5)\phm{\/}&1061.697&1.140&30\nl
10$-$0 P(5)\phm{\/}&996.181&1.112&35\nl
W 0$-$0 Q(5)\phm{\/}&1017.831&1.388&36\nl
\enddata
\tablenotetext{a}{All transitions are in the
2p$\sigma\,B\,^1\Sigma_u^+\leftarrow {\rm X}\,^1\Sigma_g^+$ Lyman band
system, unless preceded with a ``W'' which refers to the
2p$\pi\,C\,^1\Pi_u\leftarrow {\rm X}\,^1\Sigma_g^+$ Werner bands.}
\tablenotetext{b}{$N(v)$ for the main part of the profile was determined
using the pair of lines and the optical depth correction method of
Jenkins \protect\markcite{} (1996).  Other lines were used only to
define the low-level wings of the velocity profile.}
\tablenotetext{c}{The 1$-$0~R(0) line is very near the 2$-$0~R(3) line 2
echelle orders away.  These two lines were used to register lines out of
$J=0$ to those from $J=3$, as described in \S\protect\ref{vel_regs}.}
\tablenotetext{d}{The 4$-$0~R(0) line is very near the 5$-$0~R(4) line
in an adjacent echelle order.  These two lines were used to register
lines out of $J=0$ to those from $J=4$, as described in
\S\protect\ref{vel_regs}.}
\tablenotetext{e}{The velocity scale for this line was forced to be
consistent with the placement of the telluric O~I* (1027.431 and
1040.943\AA) and O~I** (1041.688\AA) lines at their computed
heliocentric velocities at the time of observation.  See
\S\protect\ref{vel_regs} for details.}
\tablenotetext{f}{The 5$-$0~R(0) line is very near the 4$-$0~P(1) line 3
echelle orders away.  These two lines were used to register lines out of
$J$=0 to those from $J=1$, as described in \S\protect\ref{vel_regs}.}
\tablenotetext{g}{The proximity of the 7$-$0~R(0) and W~0$-$0~Q(3) lines
to each other ($\Delta\lambda=0.13$\AA) enables an accurate registration
of the velocities of the lines out of $J=0$ to those out of $J=3$.}
\tablenotetext{h}{The 0$-$0~P(1) and 0$-$0~R(2) features are on top of
each other ($\Delta\lambda=0.06$\AA), but a comparison of the left-hand
component for the former with the right-hand one of the latter allows us
to link the velocity scales of absorptions out of $J=1$ and 2.  Neither
of these profiles were used in constructing the composite profiles.}
\tablenotetext{i}{Not used in building a composite profile.}
\tablenotetext{j}{The 2$-$0~R(1) line is very near the 3$-$0~R(3) line 2
orders away.  These two lines were used to register lines out of $J=1$
to those from $J=3$, as described in \S\protect\ref{vel_regs}.}
\tablenotetext{k}{The 5$-$0 R(2) line is very near the 5$-$0 P(3) line
in an adjacent order.  These two lines were used to register lines out
of $J=2$ to those with $J=3$, as described in \S\protect\ref{vel_regs}.}
\tablenotetext{l}{The W 0$-$0 Q(2) line is very near the 8$-$0 R(3) line
in an adjacent order.  These two lines were used to register lines out
of $J=2$ with those from $J=3$, as described in
\S\protect\ref{vel_regs}.} 
\tablenotetext{m}{This line was observed twice.}
\end{deluxetable}
\clearpage

\section{Results}\label{results}
\subsection{Qualitative Conclusions}\label{qualitative}

\begin{figure}
\epsscale{.8}
\plotone{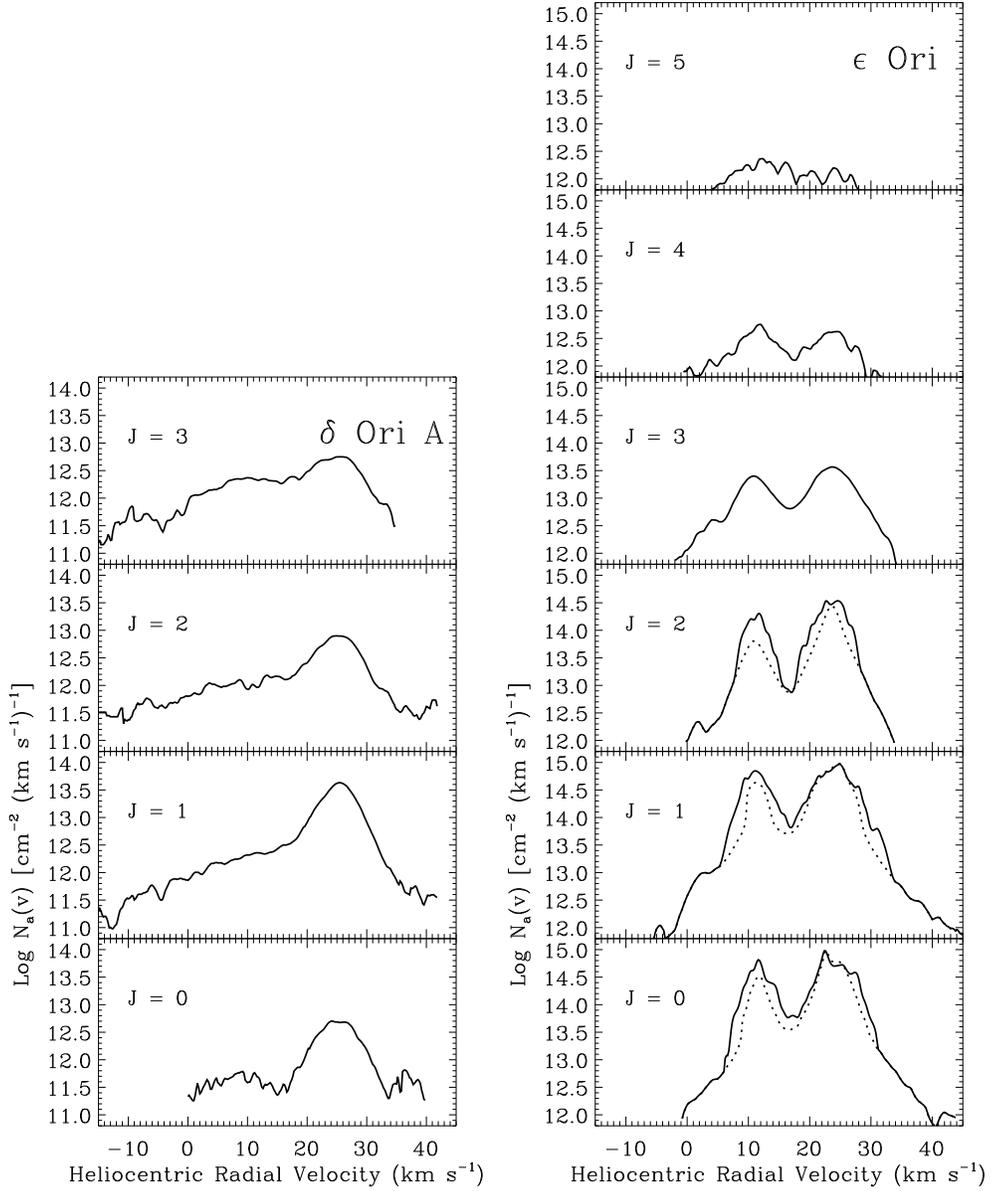}
\caption{Semilog plots of the apparent differential column densities
$N_a(v)$ (defined in Eq.~\protect\ref{N_a}) for each $J$ level of H$_2$
in the direction of $\delta$~Ori~A (left column of panels) and
$\epsilon$~Ori (right column).  For $J=0$, 1 and 2 toward
$\epsilon$~Ori, we evaluated the strong, central portions of the
profiles from the two weakest transitions using a procedure described by
Jenkins  (1996) for correcting for saturated, unresolved features inside
the main peaks.  The dotted lines show the outcome for the weakest line
without this correction.\label{naplts}}
\end{figure}

Figure~\ref{naplts} shows logarithmic depictions of the apparent column
densities (from Eq.~\ref{N_a}) for all observable $J$ levels for the two
stars. It is clear from the shapes of the profiles in Fig.~\ref{naplts}
that our results agree qualitatively with the conclusions derived by
Spitzer et al.  (1974) from {\it Copernicus\/} data, refined by a more
meticulous study by Spitzer \& Morton  (1976) in the case of
$\delta$~Ori, that the general structures seen toward both $\delta$ and
$\epsilon$ show some broadening with increasing $J$.  For instance, the
main peak in the spectrum of $\delta$~Ori~A centered at $25\,{\rm
km~s}^{-1}$ is accompanied by a shoulder at more negative radial
velocities that becomes more prominent with increasing $J$.  The shape
of the peak itself is invariant with $J$ however [measured widths (FWHM)
of linear representations of the profiles shown in Fig.~\ref{naplts} are
7, 6, 7 and $8.5\,{\rm km~s}^{-1}$ for $J=0-3$, respectively, with the
higher result for $J=3$ probably being attributable to the stronger
contribution from the shoulder].  The broadening with increasing $J$ is
less clear for $\epsilon$~Ori.  The most conspicuous effect here is the
decrease in the apparent depth of the gap between the two peaks when
going from $J=0,1,2$ to $J=3,4,5$.  For the H$_2$ lines in the spectrum
of $\epsilon$~Ori recorded by {\it Copernicus}, the two components are
not resolved, but appear as a single peak with a slight asymmetry 
(Shull 1979).

\placefigure{naplts}

The three Orion Belt stars seem to have in common a main peak centered
at a velocity of approximately $25\,{\rm km~s}^{-1}$.  It is strongest
toward $\epsilon$, of intermediate strength for $\zeta$ [compare our
Fig.~\ref{naplts} with Figs.~$2-6$ of Jenkins \& Peimbert  (1997)], and
weakest in the spectrum of $\delta$.  Maps of 21-cm emission over a
velocity range +18 to $+26\,{\rm km~s}^{-1}$ (+1 to $+9\,{\rm
km~s}^{-1}$ LSR) show a pronounced north-south ridge of atomic hydrogen
to the east of $\zeta$~Ori~A  (Chromey, Elmegreen, \& Elmegreen 1989;
Green 1991).  If the main H$_2$ peak were closely coupled with this
atomic hydrogen, we would have expected $\zeta$~Ori to show the largest
column density, $\epsilon$ an intermediate value, and $\delta$ the
smallest.  Instead, $\epsilon$ actually shows the most, with $\zeta$
being second and $\delta$ third.

At $+11\,{\rm km~s}^{-1}$ $\epsilon$ shows a peak that is almost as
strong as its main peak at $+25\,{\rm km~s}^{-1}$, while for $\delta$
this peak is weaker and spread over a large range of velocities.  H$_2$
at this velocity shows up very weakly in the spectrum of $\zeta$
[identified as Component~2 by Jenkins \& Peimbert  (1997)].  This
behavior seems to agree with the 21-cm results: an isolated cloud of
hydrogen, about 1 degree in diameter, is seen at the approximate
location of $\epsilon$ in the maps of Chromey, et al.  (1989) and Green
\& Padman  (1993) at $v_{\rm LSR}=-3$ to $-8\,{\rm km~s}^{-1}$ (i.e.,
consistent with our heliocentric velocity of $+11\,{\rm km~s}^{-1}$). 
Finally, the existence of a component at about $-1\,{\rm km~s}^{-1}$
seems to be unique to $\zeta$, but we could not identify in the 21-cm
maps any distinctive atomic hydrogen cloud that was linked to it.

While the H$_2$ profiles toward $\delta$ and $\epsilon$~Ori show some
changes in shape as the $J$ values progress to higher values, there
seems to be no manifestation of the distinctive broadening and shifts
toward more negative velocities exhibited by the most negative velocity
component [Component~1 of Jenkins \& Peimbert  (1997)] seen in the
direction of $\zeta$~Ori~A.  Our original aim in this investigation was
to study possible new examples of this phenomenon, but none were found. 
This conclusion is interesting in view of fact that these stars are
separated by only about 10~pc in projection on the sky.  As long as we
are not being misled by the existence of a considerably larger
separation longitudinally, our initial expectation was that the stellar
winds from these stars probably create a single interstellar bubble that
is common to all of them  (Castor, McCray, \& Weaver 1975; Weaver et al.
1977).  Other factors being equal, it was natural to expect that the
interactions of these winds with the ambient medium should be similar in
all three cases.

\subsection{Distribution of H$_2$ over Different $J$
Levels}\label{distribution}

The logarithms of the column densities $\int N_a(v)dv$ over different
velocity intervals and their errors are listed in Tables~\ref{Ns_delta}
and \ref{Ns_epsilon}, along with results reported by Spitzer et al. 
(1974).  For all $J$ levels in both stars, the two investigations agree
within their combined errors.  The atomic hydrogen column densities
toward these stars are given in Table~\ref{coord}.  The fractions of H
atoms bound into molecular form $2N({\rm H}_2)/[2N({\rm H}_2)+N({\rm
H~I})]$ are $7\times 10^{-6}$ for $\delta$ and $1.3\times 10^{-4}$ for
$\epsilon$.

\placetable{Ns_delta}
\placetable{Ns_epsilon}
\begin{deluxetable}{
c     
c     
c     
c     
c     
c     
c     
c     
}
\tablecolumns{8}
\tablewidth{0pt}
\tablecaption{Logarithms of Column Densities of H$_2$ toward
$\delta$~Ori~A\label{Ns_delta}}
\tablehead{
\colhead{} & \multicolumn{4}{c}{This paper} & \colhead{} &
\multicolumn{2}{c}{Spitzer et al.  (1974)} \\
\cline{2-5} \cline{7-8} \\
\colhead{} & \multicolumn{3}{c}{$\log \int N_a(v)dv$} \\
\cline{2-4} \\
\colhead{} & \colhead{Shoulder} & \colhead{Main Peak} & \colhead{Total}
\\
             \colhead{$J$}
           & \colhead{($-10$ to 18)\tablenotemark{a}}
           & \colhead{(18 to 35)\tablenotemark{a}}
           & \colhead{($-10$ to 35)\tablenotemark{a}}
           & \colhead{error\tablenotemark{b}}
           & \colhead{}
           & \colhead{$\log N$}
           & \colhead{error} \\
}
\startdata
0\dotfill&12.84&13.60&13.67&0.05&&13.54&$0.04-0.09$\nl
1\dotfill&13.60&14.44&14.50&0.05&&14.41&$0.04-0.09$\nl
2\dotfill&13.38&13.81&13.95&0.05&&13.87&$0.04-0.09$\nl
3\dotfill&13.58&13.73&13.96&0.05&&13.97&$0.04-0.09$\nl
4\dotfill&\nodata&\nodata&\nodata&\nodata&&$\leq 13.02$&\nodata\nl
5\dotfill&\nodata&\nodata&\nodata&\nodata&&$< 12.76$&\nodata\nl
Total&14.04&14.64&14.74&0.05&&14.66&$0.04-0.09$\nl
\enddata
\tablenotetext{a}{Heliocentric velocity interval for integration (${\rm
km~s}^{-1}$)}
\tablenotetext{b}{Applies to the preceding three columns.}
\end{deluxetable}
\begin{deluxetable}{
c     
c     
c     
c     
c     
c     
c     
c     
}
\tablecolumns{8}
\tablewidth{0pt}
\tablecaption{Logarithms of Column Densities of H$_2$ toward
$\epsilon$~Ori\label{Ns_epsilon}}
\tablehead{
\colhead{} & \multicolumn{4}{c}{This paper} & \colhead{} &
\multicolumn{2}{c}{Spitzer et al.  (1974)} \\
\cline{2-5} \cline{7-8} \\
\colhead{} & \multicolumn{3}{c}{$\log \int N_a(v)dv$} \\
\cline{2-4} \\
\colhead{} & \colhead{Component 1} & \colhead{Component 2} &
\colhead{Total} \\
             \colhead{$J$}
           & \colhead{(0 to 17)\tablenotemark{a}}
           & \colhead{(17 to 32)\tablenotemark{a}}
           & \colhead{(0 to 32)\tablenotemark{a}}
           & \colhead{error\tablenotemark{b}}
           & \colhead{}
           & \colhead{$\log N$} & \colhead{error} \\
}
\startdata
0\dotfill&15.38\tablenotemark{c}&15.65\tablenotemark{c}&
15.84\tablenotemark{c}&0.20&&16.08&$0.10-0.19$\nl
1\dotfill&15.53\tablenotemark{c}&15.75\tablenotemark{c}&
15.96\tablenotemark{c}&0.20&&16.36&$0.20-0.29$\nl
2\dotfill&14.90\tablenotemark{c}&15.27\tablenotemark{c}&
15.42\tablenotemark{c}&0.20&&15.34&$0.10-0.19$\nl
3\dotfill&14.22&14.42&14.64&0.05&&14.51&$0.04-0.09$\nl
4\dotfill&13.59&13.52&13.86&0.15&&13.82&$0.04-0.09$\nl
5\dotfill&13.29&13.10&13.50&$^{+0.3}_{-\infty}$&&$\leq
13.98$\tablenotemark{d}&\nodata\nl
Total&15.83&16.09&16.28&0.20&&16.57&$0.20-0.29$\nl
\enddata
\tablenotetext{a}{Heliocentric velocity interval for integration (${\rm
km~s}^{-1}$)}
\tablenotetext{b}{Applies to the preceding three columns.}
\tablenotetext{c}{Includes a correction for unresolved saturated
absorptions, using the method of Jenkins  (1996).}
\tablenotetext{d}{A value $\log N_5=13.32$ is given by Shull  (1979).}
\end{deluxetable}

\begin{figure}
\epsscale{1.0}
\plotone{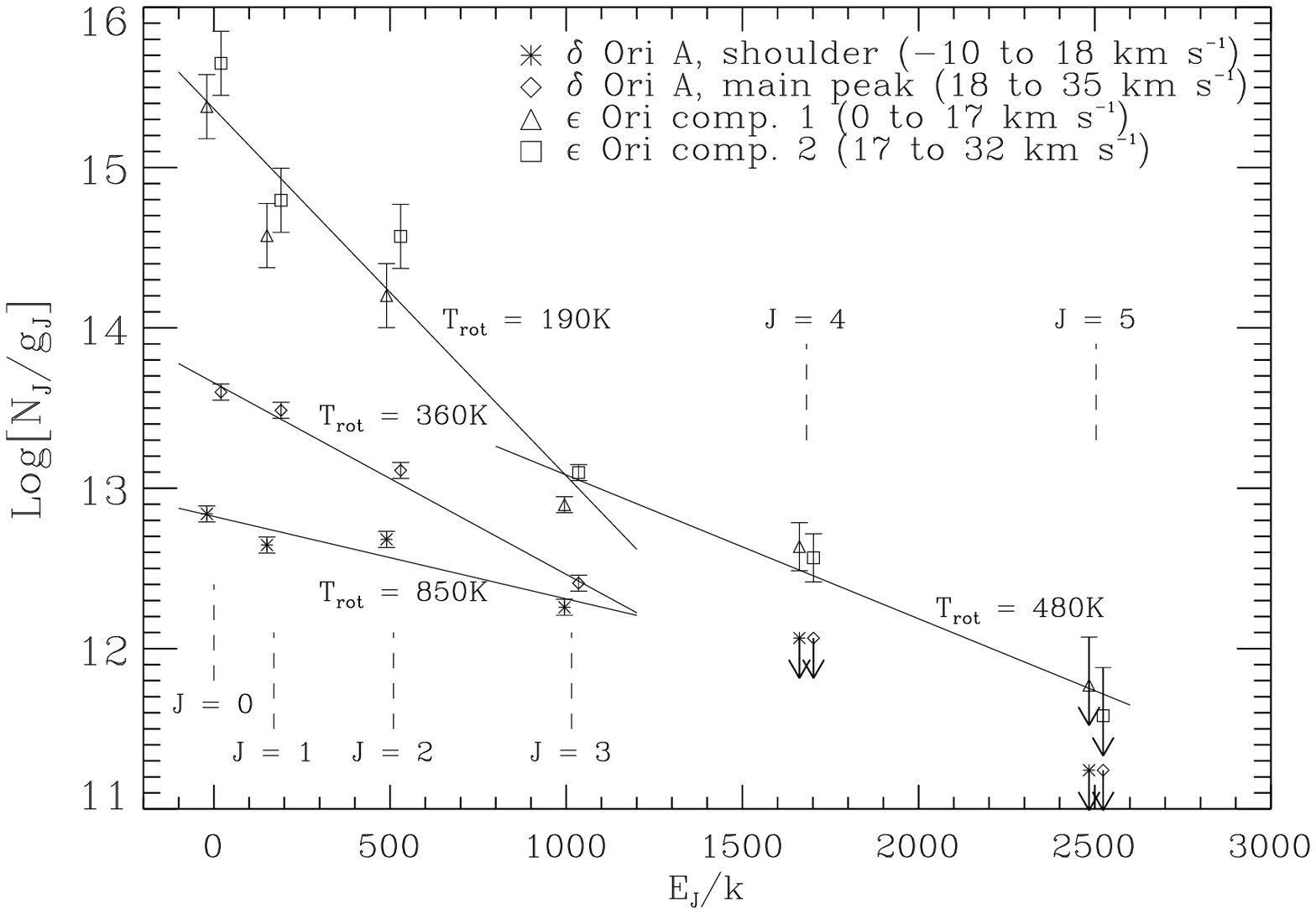}
\caption{The relationships between the column densities in different $J$
levels, $N_J$, divided by their respective statistical weights, $g_J$
(expressed in terms of the logarithm of the value in ${\rm cm}^{-2}$)
plotted against their energies $E_J/k$ (expressed in K).  The negative
inverse slopes of the best fit lines indicate the rotational
temperatures $T_{\rm rot}$.  The fits for $\epsilon$~Ori are to an
average of both velocity components, since their behaviors are so
similar.\label{h2temps}}
\end{figure}

Figure~\ref{h2temps} shows a plot of $\log(N_J/g_J)~vs.~E_J/k$ for
different components of the H$_2$ absorption observed toward $\delta$
and $\epsilon$~Ori, where $N_J$ is the column density in level $J$,
$g_J$ is the degeneracy of this level, and $E_J$ is its energy above the
$J=0$ level.  One can use the Boltzmann equation to define an effective
rotational temperature $T_{\rm rot}$ between two levels $J$ and
$J^\prime$
\begin{equation}\label{T_rot}
T_{\rm rot}={E_{J^\prime}-E_J\over
k\ln(10)[\log(N_J/g_J)-\log(N_{J^\prime}/g_{J^\prime})]}
\end{equation}
which leads to the fact that $T_{\rm rot}$ is proportional to the
negative inverse of the slope of a line joining the points for the
respective levels in Fig.~\ref{h2temps}.

\placefigure{h2temps}

For $\epsilon$~Ori, both components show nearly identical relative
populations, but they seem to exhibit two values for $T_{\rm rot}$.  One
applies to the interval $J=0-3$ and another is appropriate for $J=3-5$. 
We have determined each by finding least-squares, linear fits to the
points in Fig.~\ref{h2temps}, and the outcomes are stated near these
trend lines.  The abrupt change in slope usually signifies a transition
from temperatures that are coupled to collisions at low $J$ to ones that
are dominated by optical pumping at higher $J$.  Another effect that may
contribute to the relationship at low $J$ is a diminution of the pumping
rate at the lowest $J$ levels caused by the pumping lines becoming
optically deep.

The radiative decay rates at large $J$ are so fast that collisions have
a practically negligible effect.  For this reason, to a reasonable
approximation we can use the ratios $N_4/N_3$ and $N_5/N_3$ as
indicators of the density of pumping radiation (when the radiation
density increases, a large proportion of molecules in $J=3$ are pumped
to the upper electronic level before they have a chance to decay to
$J=1$ -- see \S V(a) of Spitzer \& Morton  (1976) for a more complete
discussion and representative examples).  When we compare our ratios for
$\epsilon$~Ori with models presented in Table~5 of Spitzer \& Morton 
(1976), we conclude that the pumping rate $\beta=3-10\times 10^{-9}{\rm
s}^{-1}$ for Component~1 and $1-3\times 10^{-9}{\rm s}^{-1}$ for
Component~2.  For comparison, Jura  (1974) calculated that
$\beta=5\times 10^{-10}{\rm s}^{-1}$ in the general vicinity of the Sun. 
The elevation in $\beta$ for H$_2$ in front of $\epsilon$~Ori seems
consistent with what one would expect for gas that is in the general
vicinity of the association but not especially close to one of the
bright, early-type stars.

We have less complete information on the distribution with $J$ for
$\delta$~Ori~A because the overall column density of H$_2$ is lower. 
Since our values for $N_J$ are nearly the same as those of Spitzer \&
Morton  (1976), we adopt their interpretation for the conditions that
lead to the observed relative populations.  Spitzer \& Morton found that
for their Component~3 of H$_2$ for $\delta$~Ori~A, which approximately
corresponds to our main peak, $\beta=2\times 10^{-10}{\rm s}^{-1}$ for
$J=0$ and 1, but $\beta=5\times 10^{-10}{\rm s}^{-1}$ for $J=2$.  This
seems to indicate that the gas in the main component is rather distant
from the Orion stars.  The lower pumping rate in the lowest $J$ levels
probably arises from the diminution of flux from distant stars by
molecular complexes unrelated to the one seen toward $\delta$~Ori~A,
since in our spectrum only the strongest lines, such as those the Werner
1$-$0~Q(1) and 2$-$0~Q(1) transitions, have central optical depths
appreciably greater than about 2.  The suggested remoteness for this
component of H$_2$ from the Orion association is an unexpected
conclusion when one considers the fact that its velocity is very similar
to one of the main components in front of $\epsilon$~Ori (and also
$\zeta$~Ori~A).  Perhaps the molecules in front of $\delta$~Ori~A are
heavily shielded from the pumping radiation by other parts of the same
gas complex in the vicinity of the Orion association.

For the shoulder component toward $\delta$~Ori~A, the upper limit for
$J_5/J_3<0.15$ found by combining our results with those of Spitzer et
al.  (1974) indicates that $\beta < 4\times 10^{-9}{\rm s}^{-1}$.  In
the limit of a low density, the high rotation temperature $T_{\rm
rot}=850$K for $J=0-3$ would be consistent with a considerably larger
value of $\beta$, so perhaps the local kinetic temperature is about 850K
and the density is large enough to have collisions couple $T_{\rm rot}$
to this temperature.

\section{Interpretation}\label{interpret}

We turn now to the question of what we can learn from the absence of the
broadening and velocity shifts in the spectra of $\delta$ and $\epsilon$
Ori that Jenkins \& Peimbert  (1997) found in the most negative velocity
H$_2$ feature for $\zeta$~Ori~A.  In particular, are there other
differences between the two stars featured here and $\zeta$~Ori~A that
are noteworthy and might give us further insights on the phenomenon? 
Does this information support or refute the interpretation that it
arises from H$_2$ forming in a cooling zone behind a bow shock, one that
could be created by the collision of a gas flow coming toward us and a
foreground obstruction, as suggested by Jenkins \& Peimbert  (1997)?

In addressing the question just posed, we examined two types of
evidence.  First, we searched for differences revealed by other spectral
features for these stars (\S\ref{hi_vel}).  Second, we looked for
relationships of the stars with structural features of the interstellar
medium detected by other means (\S\ref{structures}).  In both cases, we
noted certain properties that set $\zeta$~Ori~A apart from $\delta$ and
$\epsilon$~Ori.

\subsection{Singly Ionized Gas at High Negative
Velocities}\label{hi_vel}

\begin{figure}
\epsscale{1.0}
\plotone{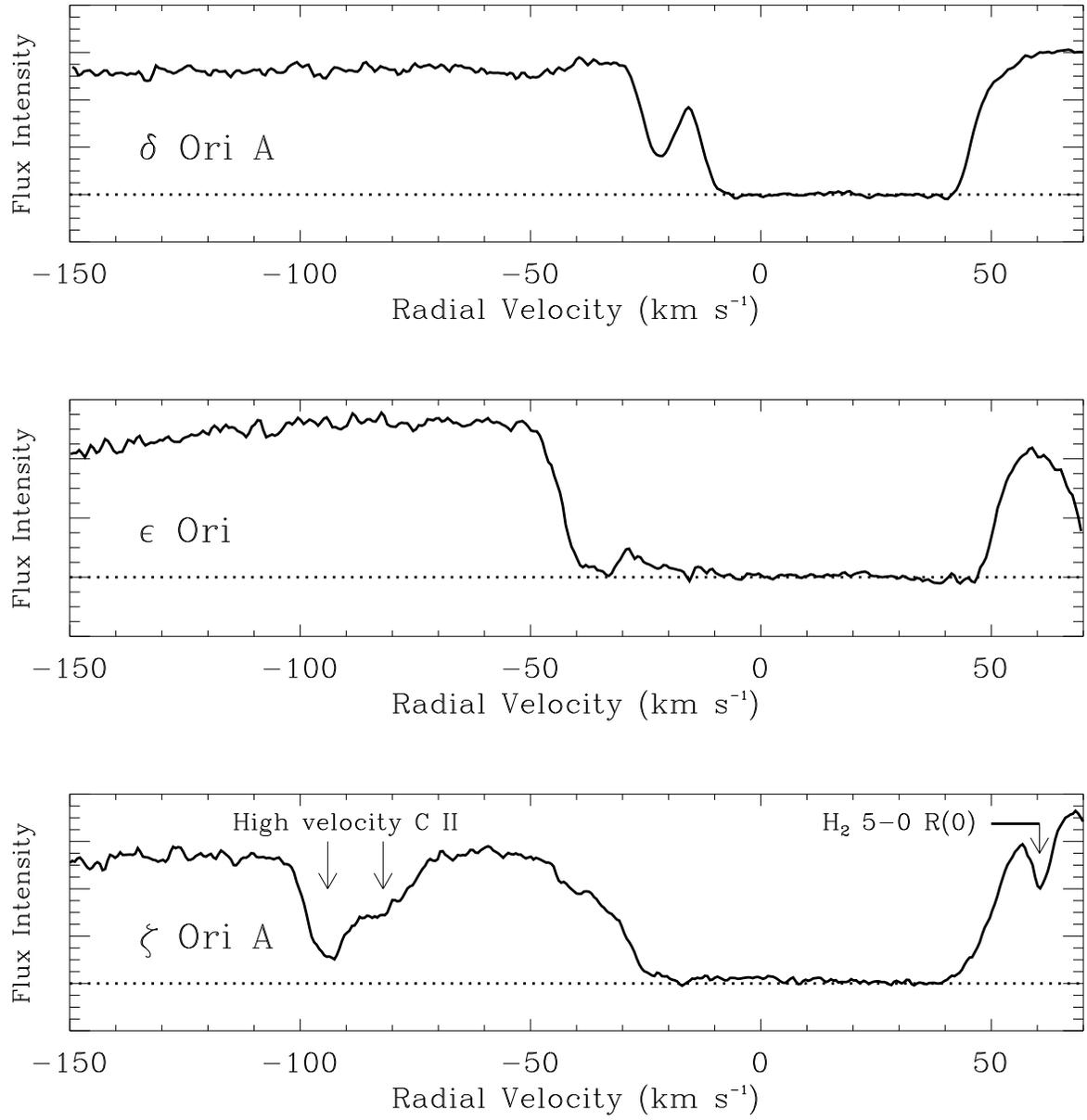}
\caption{Fluxes in the vicinity of the C~II transition at 1036.337\AA\
for $\delta$, $\epsilon$ and $\zeta$~Ori, depicted on a scale of
heliocentric radial velocities for this transition.\label{ciispec}}
\end{figure}

\begin{figure}
\epsscale{1.0}
\plotone{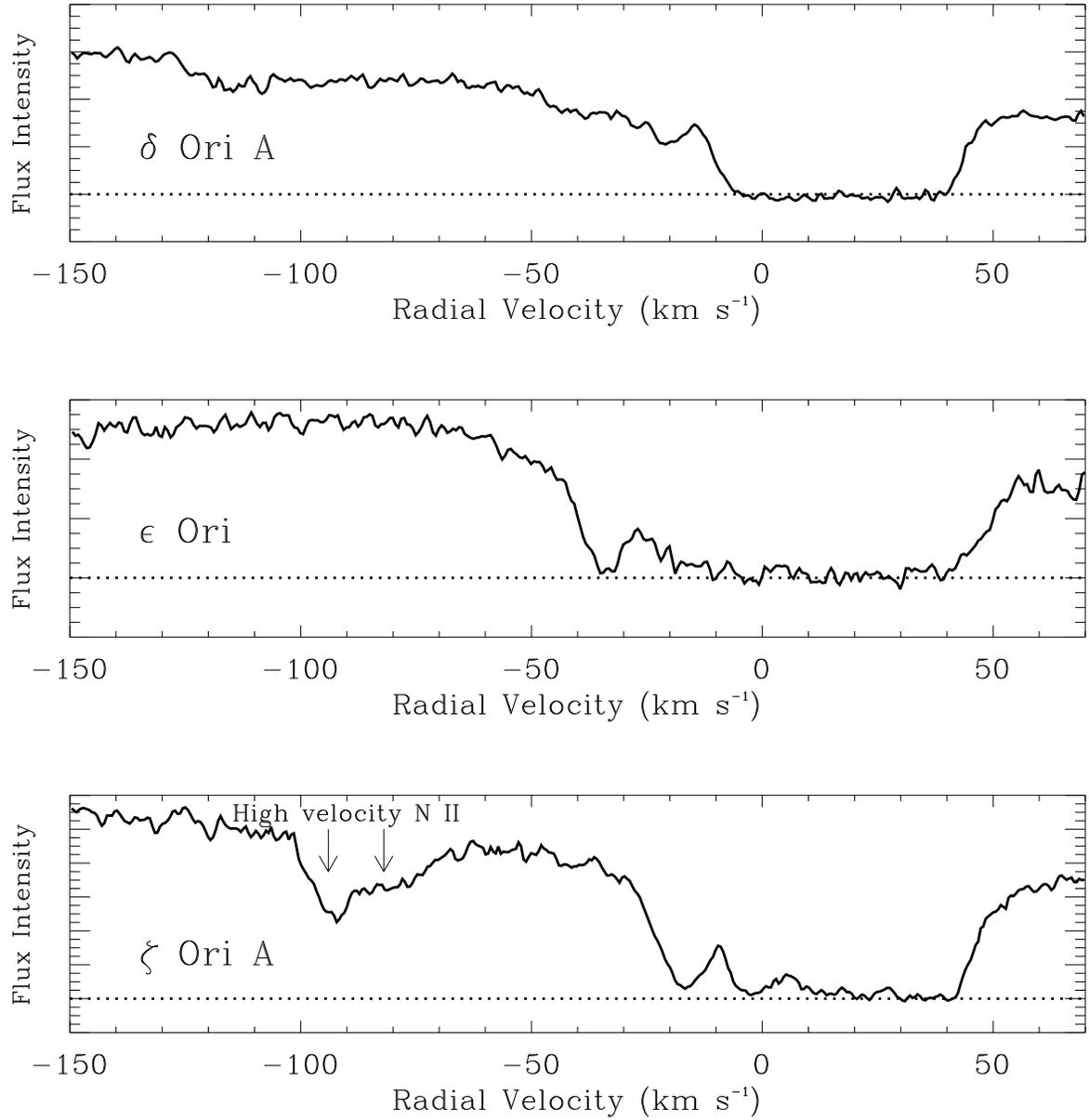}
\caption{Plots similar to those of Fig.~\protect\ref{ciispec} that apply
to the N~II transition at 1083.990\AA.\label{niispec}}
\end{figure}

Figures~\ref{ciispec} and \ref{niispec} show IMAPS recordings of the
strong resonance lines of the singly ionized forms of C and N,
respectively, for the three Orion Belt stars.  The most striking
difference that sets $\zeta$ apart from $\delta$ and $\epsilon$ is the
absorption from material that is moving at a radial velocity that ranges
between $-100$ and $-70\,{\rm km~s}^{-1}$.  In the spectrum of
$\zeta$~Ori~A, similar components at these velocities are seen in C~III,
N~III, Si~III and S~III  (Cowie, Songaila, \& York 1979) and Al~III
(medium-resolution GHRS spectrum in the {\it HST}
archive\footnote{Exposure identification: Z165040DM.}), but are not
visible in the strong transitions of O~I and N~I.  The interpretation of
Jenkins \& Peimbert  (1997) regards this as the upstream material that
feeds the shock front that is responsible for the H$_2$ component at
$-1\,{\rm km~s}^{-1}$ which shows the unusual behavior with increasing
$J$.  The correlation of the presence or absence of the high velocity
gas with the phenomenon noted by Jenkins \& Peimbert seems to favor
their shock interpretation over completely unrelated alternatives.

\placefigure{ciispec}

\placefigure{niispec}

\subsection{Interstellar Gas Structures}\label{structures}

\begin{figure}
\epsscale{1.0}
\plotone{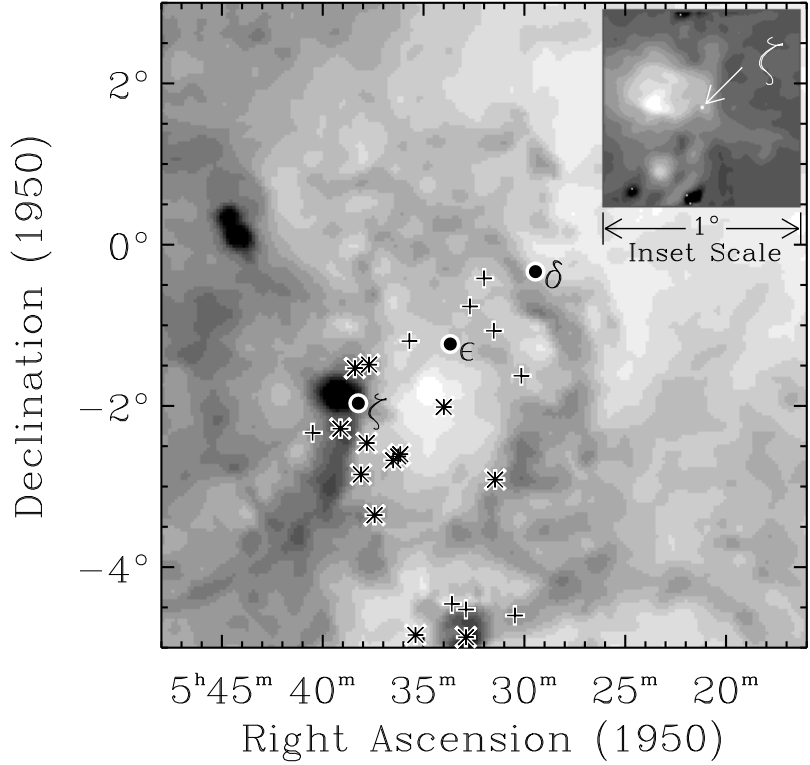}
\caption{Locations of $\delta$, $\epsilon$ and $\zeta$~Ori with respect
to diffuse emission by dust at $100\mu$m recorded by IRAS, but cleaned
by Schlegel, Finkbeiner \& Davis  (1998) and then renormalized to
represent $E(B-V)$, using information about grain temperatures derived
from DIRBE observations at longer wavelengths.  The darknesses in this
depiction are proportional to the logarithm of this computed color
excess.  The dense cloud near $\zeta$ is Lynds~1630.  Values of the
computed $E(B-V)$ at the positions of $\delta$, $\epsilon$ and $\zeta$
in this map equal 0.25, 0.30 and 5.14, respectively.  For $\delta$ and
$\epsilon$~Ori, their actual values of $E(B-V)$ of 0.09 and 0.05
indicate that more than half of the dust emission originates from
material behind the stars.  The locations of B-type stars whose IUE
spectra were examined by Shore  (1982) and Sonneborn, et al.  (1988) are
indicated: those with UV absorption features at velocities equal to or
more negative than $-50\,{\rm km~s}^{-1}$ are shown with asterisks,
while those without such features are shown with crosses.  The inset
shows a HIRES processing (beam width $<$ 1\arcmin\ FWHM) of the 100$\mu$
IRAS flux in the immediate vicinity of $\zeta$~Ori.\label{dust}}
\end{figure}

The interstellar material in the general vicinity of the Orion OB1
association contains a number of well-studied gas complexes and
prominent emission regions [see Goudis  (1982) for a detailed summary].
The lines of sight to $\epsilon$ and $\delta$ seem unremarkable, but
that toward $\zeta$~Ori~A points in a direction that is very close to
the densest concentration of material within the Orion~B molecular cloud
complex.  This cloud has a large visual extinction and appears as entry
number 1630 in the catalog of dark nebulae compiled by Lynds  (1962). 
It has a significant internal density, and it contains a large number of
very young, but highly obscured stars   (Strom, Strom, \& Vrba 1976). 
The striking difference between the direction toward $\zeta$ and those
toward the other two stars is shown in a map of thermal emission from
dust  (Schlegel, Finkbeiner, \& Davis 1998) recorded by the IRAS
satellite, depicted in Fig.~\ref{dust}, and likewise in maps of CO
emission  (Kutner et al. 1977; Maddalena et al. 1986).  Maps of 21-cm
emission in this region of Orion show no indication of this strong
contrast  (Chromey, Elmegreen, \& Elmegreen 1989; Green 1991; Green \&
Padman 1993), probably because the emission from this cold, highly
condensed cloud is very badly saturated.

\placefigure{dust}
\begin{figure}
\plotone{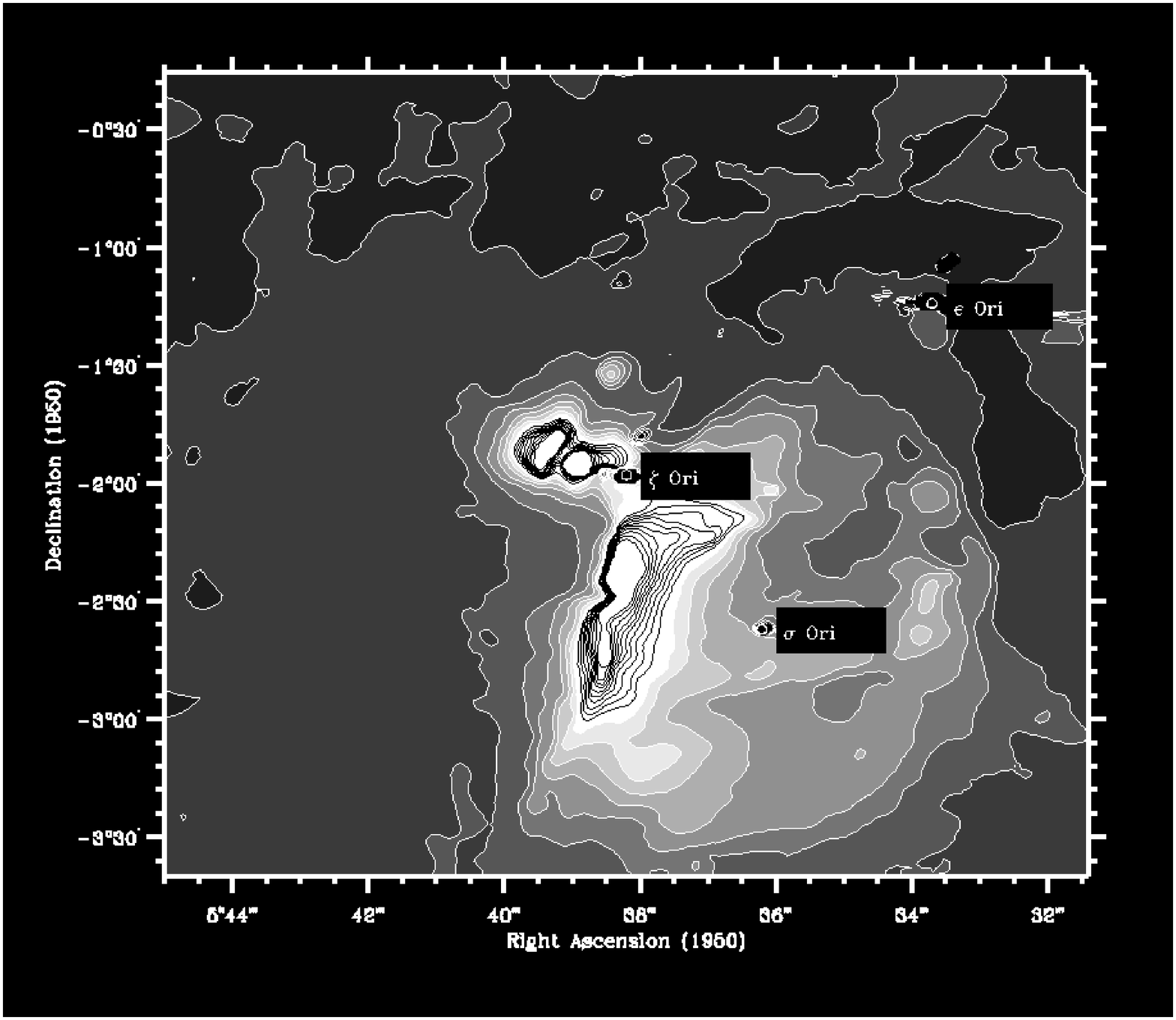}
\caption{A contour map of the intensity of H$\alpha$ emission in the
general vicinity of the dense cloud Lynds~1630, recorded by Gaustad et
al.  (1999).  The bright north-south ridge of emission is IC~434, and
the region of high obscuration in Lynds~1630 is just to the east of it. 
The highest contour levels have been clipped.  Fluxes depicted here are
only relative, since the sensitivity to H$\alpha$ emission and the zero
offset caused by geocoronal H$\alpha$ emission have not yet been
determined.\label{gaustad}}
\end{figure}

Could Lynds~1630 and its bright edge IC~434 (see Fig.~\ref{gaustad}) on
the side toward $\zeta$ and $\sigma$~Ori  (Pottasch 1956) be the
obstruction proposed by Jenkins \& Peimbert  (1997) that is needed to
create a stationary bow shock in the stellar wind material?   The
plausibility of this idea very much depends on whether or not Lynds~1630
is in front of or behind $\zeta$~Ori.

At first glance, one might suppose that the cloud must be in front,
since the main body of the cloud appears much darker in H$\alpha$
emission than the edge toward $\zeta$~Ori.  Also, the existence of the
Horsehead nebula as a dark silhouette against the H$\alpha$ emission is
an additional clue, although there is a reasonable chance that the
background excitation at this particular location could arise from
radiation from $\sigma$~Ori (which might be more distant) rather than
$\zeta$~Ori. The bright, foreground H~II region NGC~2024 [also known as
the Orion~B radio source  (Barnes et al. 1989; van der Werf et al.
1993)] is powered by UV radiation from one or more highly obscured
star(s) embedded within it, not by $\zeta$~Ori  (Grasdelen 1974).

On closer inspection, however, we conclude that the indications offered
by the structure of the most easily evident H$\alpha$ emission could be
misleading.  If we disregard the very bright patch of H$\alpha$ emission
associated with the edges of the Lynds~1630 cloud and consider the
fainter, more broadly distributed emission in the general region, a
different picture emerges.  Nearly all of the volume inside Barnard's
Loop  (Pickering 1890; Barnard 1894) is filled with a faint glow created
by the ionizing radiation from the Orion association  (Sivan 1974). 
Fig.~\ref{gaustad} shows a contour map of H$\alpha$ emission in the
vicinity of $\zeta$ and $\epsilon$~Ori, well inside the perimeter of
Barnard's Loop.  This map was constructed from the survey of Gaustad et
al. (1999)\footnote{Preliminary data in digital form were provided by
Dr. J.~Gaustad in advance of publication of the actual results.}, and it
clearly shows that the brightness level at the locations where there is
significant obscuration from the cloud (but somewhat removed from
NGC~2024) is comparable to the flux from directions approximately midway
between $\zeta$ and $\epsilon$~Ori, well beyond the cloud's edge to the
northwest. [An earlier study of H$\alpha$ emission in this vicinity by
Vidal  (1980) also shows this result, but not as clearly.]  On the one
hand, this finding suggests that the Lynds~1630 cloud is behind the Belt
Stars, if they were responsible for ionizing hydrogen and producing
H$\alpha$ emission in front of the cloud or right on the cloud's front
edge.  On the other hand, we could argue that one or more stars other
than $\zeta$~Ori, such as $\sigma$~Ori (spectral type O9.5$\,$V), are
responsible for most of the emission and are closer to us than
$\zeta$~Ori.  Unfortunately, spectroscopic or Hipparcos parallaxes are
not accurate enough to resolve this issue.  If, by contrast, we had
found that the H$\alpha$ emission were darker in front of Lynds~1630
than in directions close to $\epsilon$~Ori, we might have had a more
positive indication that the cloud was well in front of the Belt Stars
and appears as a silhouette.

\placefigure{gaustad}

Other evidence that disfavors Lynds~1630 being in the foreground, but
again not conclusively, arises from the distribution of dust shown in
Fig.~\ref{dust}.  The amount of infrared 100$\mu$ emission from the
cloud's dust in the exact direction of $\zeta$~Ori leads to an expected
B$-$V color excess that seems to be far greater than the star's actual
$E(B-V)$~=~0.05.  A HIRES processing\footnote{The HIRES processing was
performed at the Infrared Processing and Analysis Center, using the
``Maximum Correlation Method'' to create images with a high angular
resolution  (Aumann, Fowler, \& Melnyk 1990).} of the 100$\mu$ image
(see inset of Fig.~\ref{dust}) shows that we are not being misled by the
smoothing in the representation generated by Schlegel, et al.  (1998)
combined with the star's proximity to a sharp edge of the cloud. 
Nevertheless, the gas could be clumpy on scales even smaller than the
HIRES picture, and the line of sight to $\zeta$~Ori could miss the
clumps.

The indications that Lynds~1630 may be behind $\zeta$~Ori cast doubt on
the interpretation that it is a conspicuous barrier for the stellar wind
from the star.  Nevertheless, we still should not overlook the
possibility that its presence in the same general direction of the star
may still be significant.  The cloud contains very dense concentrations
of molecular material, as revealed by high-resolution maps of molecular
emission by Lada, Bally \& Stark  (1991) (CS), Bally, Langer \& Liu 
(1991) ($^{13}{\rm CO}$), and Stark \& Bally  (1982) ($^{12}{\rm CO}$). 
It is often true that clouds containing compact molecular regions are
sites of star formation, and the Orion~B cloud is no exception  (Bence
\& Padman 1996; Chernin 1996; Launhardt et al. 1996; Zinnecker,
McCaughrean, \& Rayner 1996; Sandell et al. 1999).  One product of star
formation is the creation of high speed, bipolar flows of gas  (Bally
1982; Genzel \& Downes 1982; K\"onigl 1996) with velocities that have
about the same magnitude as the component in front of $\zeta$~Ori~A
shown in \S\ref{hi_vel}  (Eisl\"offel 1996; Hirth, Mundt, \& Solf 1996). 
If such flows penetrate beyond the edge of the dense cloud and into a
low-density medium, they could extend over large distances  (Reipurth,
Bally, \& Devine 1997), perhaps even farther than the remote, multiple
shocks  (Bence \& Padman 1996) that can be detected in emission.  Still
another possibility is that the high speed gas arises from the breakout
of material from an embedded, recently formed H~II region  (Tenorio-Tagle 1979; Sandford, Whitaker, \& Klein 1982; Subrahmanyan et al.
1997).

In \S\ref{hi_vel} we pointed out that $\delta$ and $\epsilon$~Ori do not
show gas at high negative velocities, while $\zeta$~Ori does.  We can
investigate this phenomenon further by examining the outcome of a survey
of UV absorption features in B-type stars carried out by Shore  (1982)
and Sonneborn, et al.  (1988).  In Fig.~\ref{dust} we indicate the
positions of these stars and with different symbols show whether they do
(*) or do not (+) exhibit features from singly-charged ions at
velocities equal to or more negative than $-50\,{\rm km~s}^{-1}$.  It
seems clear that stars in the vicinity of Lynds~1630 usually show the
high-velocity features, while those that are somewhat removed in a
lateral direction, such as the targets in the general vicinity of
$\delta$ and $\epsilon$~Ori, do not.  This outcome suggests the
possibility that the dense molecular cloud may be the source of high
velocity gas.  While this may be so, we should not overlook the fact
that high velocity features are seen elsewhere in the Orion region 
(Cohn \& York 1977; Cowie, Songaila, \& York 1979), with some of them at
locations outside the coverage of Fig.~\ref{dust}.  With this in mind,
we could equally well propose that a more general, coherent expansion of
gas from a larger scale event is taking place.  In this picture, the
appearances of high velocity features in Fig.~\ref{dust} could arise
from just random fragments of enhanced density in the expanding
material.

\section{Concluding Remarks}\label{remarks}

In studying the spectra of $\delta$ and $\epsilon$~Ori, we had high
expectations of finding duplications of the unusual findings of Jenkins
\& Peimbert  (1997) for one of the H$_2$ components toward
$\zeta$~Ori~A.  They reported that the profile became broader and
shifted toward more negative radial velocities as the rotational
excitation increased from $J=0$ to 5.  Our optimism about seeing more
examples of this phenomenon was based on the premise that, in Jenkins \&
Peimbert's interpretation of a bow shock as its source, the stellar wind
that may have created this shock front was probably a composite
phenomenon common to all three stars in the Orion Belt.  The apparent
lack of this effect for $\delta$ and $\epsilon$~Ori deprives us of the
chance to see additional manifestations of it, but opens the way for us
to examine circumstantial evidence that could give us more insight on
the validity of the shock interpretation.

Aside from not showing the strange behavior of H$_2$, $\delta$ and
$\epsilon$~Ori differ from $\zeta$~Ori in two other respects that could
have a bearing on the shock interpretation.  First, $\zeta$ shows
absorption by singly and doubly ionized species at a velocity of about
$-90\,{\rm km~s}^{-1}$, while the others do not.  This high velocity
material was a good candidate for the upstream gas that could have
created the bow shock.  Second, in projection $\zeta$~Ori is very near
the edge of the dense cloud Lynds~1630, an accumulation of gas that
exhibits a bright edge of ionized material (IC~434) on the side towards
$\zeta$~Ori and holds within it many compact regions of star formation. 
The sight lines to $\delta$ and $\epsilon$ Ori are, by contrast,
unremarkable.

The real source of the gas at $-90\,{\rm km~s}^{-1}$ that is seen in the
spectrum of $\zeta$~Ori is uncertain.  It may be some byproduct of the
winds from the stars, but it is not easy to understand why similar
components are not seen toward $\delta$ and $\epsilon$~Ori.  An
interesting alternative is the possibility that the $-90\,{\rm
km~s}^{-1}$ component has nothing to do with stellar winds from the Belt
stars.  The complex of very dense gas near the $\zeta$~Ori line of sight
is known for its plentiful star forming regions.  Such regions typically
discharge gas at high velocities, starting with flows that are highly
collimated.  Perhaps one such flow could escape from its gas envelope
and be responsible for the high velocity component that shows up for
$\zeta$~Ori in Figs.~\ref{ciispec} and \ref{niispec}.  A well-directed
flow could provide a reasonable way to explain the absence in the
spectra of $\delta$ and $\epsilon$~Ori of any feature analogous to the
$-90\,{\rm km~s}^{-1}$ component exhibited by $\zeta$~Ori~A.

New insights on the problem discussed in this paper may arise from
observations by the Far Ultraviolet Spectroscopic Explorer (FUSE), which
has the capability of examining a very much broader sample of cases in
its lifetime.  However its lower wavelength resolution will make it
harder to study the detailed structures in the H$_2$ profiles, and
productive research in this area may require the existence of even more
extreme cases of profile variations with $J$ than those shown toward
$\zeta$~Ori~A.

\acknowledgements

The observations reported in this paper were recorded by IMAPS during
its second orbital flight, which was carried out on the ORFEUS-SPAS~II
mission in 1996.  This effort was a joint undertaking of the US and
German space agencies, NASA and DARA.  The successful execution of our
observations was the product of efforts over many years by engineering
teams at Princeton University Observatory, Ball Aerospace Systems Group
(the industrial subcontractor for the IMAPS instrument) and Daimler-Benz
Aerospace (the German firm that built the ASTRO-SPAS spacecraft and
conducted mission operations).  Contributions to the success of IMAPS
also came from the generous efforts by members of the Optics Branch of
the NASA Goddard Space Flight Center (grating coatings and testing) and
from O.~H.~W.~Siegmund and S.~R.~Jelinsky at the Berkeley Space Sciences
Laboratory (deposition of the photocathode material).  We are grateful
B. T. Draine for his comments on an early draft of the paper and to J.
Gaustad for supplying the digital representation of the H$\alpha$ image
that was the basis of our Fig.~\ref{gaustad}.  This research was
supported by NASA grant NAG5$-$616 to Princeton University.

\end{document}